\documentclass[12pt,a4paper]{article}

\usepackage{amsmath}
\usepackage{amssymb}
\usepackage[nosort]{cite}
\usepackage[hyperref,bulletsep]{collect}
\usepackage{epic}
\def\gfxon{\usepackage[final]{graphicx}}

\gfxon

\newcommand{\Eqn}[1]{&\hspace{-0.5em}#1\hspace{-0.5em}&}
\newcommand{\tu}{{\tilde{u}}}
\newcommand{\tv}{{\tilde{v}}}
\newcommand{\tw}{{\tilde{w}}}
\newcommand{\tK}{{\tilde{K}}}
\newcommand{\tS}{{\tilde{H}}}
\newcommand{\dtS}{\tilde{\hspace{-.3em}\phantom{H}}
                   \tilde{\hspace{-.6em}S\hspace{-.0em}}\hspace{.0em}}
\newcommand{\sone}{{\mathit{1}}}
\newcommand{\stwo}{{\mathit{2}}}
\newcommand{\sonedot}{{\dot{\mathit{1}}}}
\newcommand{\stwodot}{{\dot{\mathit{2}}}}

\newcommand{\dualityarrow}{\Downarrow}

\newcommand{\hF}{\tilde F}
\newcommand{\cF}{\hat F}

\font\numbers=cmss12
\font\upright=cmu10 scaled\magstep1
\def\stroke{\vrule height8pt width0.4pt depth-0.1pt}
\def\topfleck{\vrule height8pt width0.5pt depth-5.9pt}
\def\botfleck{\vrule height2pt width0.5pt depth0.1pt}
\def\Zmath{\vcenter{\hbox{\numbers\rlap{\rlap{Z}\kern 0.8pt\topfleck}\kern2.2pt
                   \rlap Z\kern 6pt\botfleck\kern 1pt}}}
\def\Qmath{\vcenter{\hbox{\upright\rlap{\rlap{Q}\kern
                   3.8pt\stroke}\phantom{Q}}}}
\def\Nmath{\vcenter{\hbox{\upright\rlap{I}\kern 1.7pt N}}}
\def\Cmath{\vcenter{\hbox{\upright\rlap{\rlap{C}\kern
                   3.8pt\stroke}\phantom{C}}}}
\def\Rmath{\vcenter{\hbox{\upright\rlap{I}\kern 1.7pt R}}}
\def\bbZ{\ifmmode\Zmath\else$\Zmath$\fi}
\def\bbQ{\ifmmode\Qmath\else$\Qmath$\fi}
\def\bbN{\ifmmode\Nmath\else$\Nmath$\fi}
\def\bbC{\ifmmode\Cmath\else$\Cmath$\fi}
\def\bbR{\ifmmode\Rmath\else$\Rmath$\fi}

\newcommand{\bea}{\begin{eqnarray}}
\newcommand{\eea}{\end{eqnarray}}
\newcommand{\beq}{\begin{equation}}
\newcommand{\eeq}{\end{equation}}

\def\({\left(}
\def\){\right)}

\def\tS{\tilde S}

\makeatletter
\let\old@makecaption=\@makecaption
\def\@makecaption{\small\old@makecaption}
\makeatother

\makeatletter \@addtoreset{equation}{section} \makeatother

\makeatletter
\let\old@startsection=\@startsection
\renewcommand{\@startsection}[6]{\old@startsection{#1}{#2}{#3}{#4}{#5}{#6\mathversion{bold}}}
\makeatother

\let\oldPhi=\Phi
\let\oldOmega=\Omega
\let\oldLambda=\Lambda
\let\oldPsi=\Psi
\let\oldGamma=\Gamma
\let\oldDelta=\Delta
\let\oldSigma=\Sigma
\let\oldTheta=\Theta
\let\oldPi=\Pi
\renewcommand{\Phi}{\mathnormal{\oldPhi}}
\renewcommand{\Omega}{\mathnormal{\oldOmega}}
\renewcommand{\Psi}{\mathnormal{\oldPsi}}
\renewcommand{\Gamma}{\mathnormal{\oldGamma}}
\renewcommand{\Sigma}{\mathnormal{\oldSigma}}
\renewcommand{\Delta}{\mathnormal{\oldDelta}}
\renewcommand{\Theta}{\mathnormal{\oldTheta}}
\renewcommand{\Pi}{\mathnormal{\oldPi}}
\renewcommand{\Lambda}{\mathnormal{\oldLambda}}

\newcommand{\indup}[1]{_{\mathrm{#1}}}
\newcommand{\indups}[1]{_{\mathrm{\scriptscriptstyle #1}}}

\newcommand{\rep}[1]{{\mathbf{#1}}}
\newcommand{\matr}[2]{\left(\begin{array}{#1}#2\end{array}\right)}

\newcommand{\alg}[1]{\mathfrak{#1}}
\newcommand{\grp}[1]{\mathrm{#1}}

\newcommand{\atopfrac}[2]{\genfrac{}{}{0pt}{}{#1}{#2}}
\newcommand{\sfrac}[2]{{\textstyle\frac{#1}{#2}}}
\newcommand{\half}{\sfrac{1}{2}}

\newcommand{\pint}{\makebox[0pt][l]{\hspace{3.4pt}$-$}\int}

\newcommand{\superN}{\mathcal{N}}
\newcommand{\gym}{g\indups{YM}}
\newcommand{\fld}[1]{\mathcal{#1}}
\newcommand{\fldZ}{\fld{Z}}
\newcommand{\fldF}{\fld{F}}
\newcommand{\fldW}{\fld{W}}
\newcommand{\cder}{\fld{D}}
\newcommand{\Op}{\mathcal{O}}

\newcommand{\order}[1]{\mathcal{O}(#1)}

\newcommand{\sgrad}{\eta}

\newcommand{\Real}{\mathbb{R}}

\newcommand{\Integers}{\mathbb{Z}}

\newcommand{\lrbrk}[1]{\left(#1\right)}
\newcommand{\bigbrk}[1]{\bigl(#1\bigr)}
\newcommand{\Bigbrk}[1]{\Bigl(#1\Bigr)}

\newcommand{\comm}[2]{[#1,#2]}

\newcommand{\acomm}[2]{\{#1,#2\}}

\newcommand{\state}[1]{\mathopen|#1\mathclose\rangle}

\newcommand{\set}[1]{\{#1\}}
\newcommand{\bigset}[1]{\bigl\{#1\bigr\}}
\newcommand{\sset}[2]{\{#1||#2\}}

\newcommand{\dil}{\mathbf{E}}

\newcommand{\sheetsign}{\varepsilon}
\newcommand{\contour}[1]{\mathcal{#1}}
\newcommand{\resolvsl}[1][]{\makebox[0pt][l]{\hspace{0.06em}$/$}#1G}
\newcommand{\sheetsl}[1][]{\makebox[0pt][l]{\hspace{0.06em}$/$}#1p}

\newcommand{\nln}{\nonumber\\}
\newcommand{\nl}{\nonumber\\&\hspace{-4\arraycolsep}&\mathord{}}

\newcommand{\earel}[1]{\mathrel{}&\hspace{-2\arraycolsep}#1\hspace{-2\arraycolsep}&\mathrel{}}
\newcommand{\eq}{\earel{=}}

\makeatletter
\newcommand{\newop}[2]{\def#1{\mathop{\operator@font #2}\nolimits}}
\newcommand{\newbin}[2]{\def#1{\mathbin{\operator@font #2}}}
\makeatother

\newop{\Re}{Re}
\newop{\Im}{Im}
\newop{\diag}{diag}
\newop{\rank}{rank}
\newop{\Tr}{Tr}
\newop{\tr}{tr}
\newop{\str}{str}
\newop{\sdet}{sdet}
\newop{\sign}{sign}

\def\[{\begin{equation}}
\def\]{\end{equation}}
\def\<{\begin{eqnarray}}
\def\>{\end{eqnarray}}

\makeatletter
\def\mr@ignsp#1 {\ifx\:#1\@empty\else #1\expandafter\mr@ignsp\fi}%
\newcommand{\multiref}[1]{\begingroup
\xdef\mr@no@sparg{\expandafter\mr@ignsp#1 \: }%
\def\mr@comma{}%
\@for\mr@refs:=\mr@no@sparg\do{\mr@comma\def\mr@comma{,}\ref{\mr@refs}}%
\endgroup}
\makeatother

\newcommand{\hypref}[2]{\ifx\href\asklfhas #2\else\href{#1}{#2}\fi}

\newcommand{\secref}[1]{Sec.~\multiref{#1}}

\newcommand{\appref}[1]{App.~\multiref{#1}}

\newcommand{\figref}[1]{Fig.~\multiref{#1}}
\renewcommand{\eqref}[1]{(\multiref{#1})}

\newenvironment{bulletlist}{\begin{list}{$\bullet$}{\leftmargin1.5em\itemsep0pt}}{\end{list}}

\ifx\href\asklfhas\newcommand{\href}[2]{#2}\fi
\newcommand{\arxivno}[1]{\href{http://arxiv.org/abs/#1}{#1}}

\begin{document}

\thispagestyle{empty}
\begin{flushright}\footnotesize
\texttt{\arxivno{hep-th/0503200}}\\
\texttt{ITEP-TH-23/05}\\
\texttt{LPTENS-05/12}\\
\texttt{PUTP-2153}\\
\texttt{UUITP-05/05}\\
\vspace{0.5cm}
\textit{Dedicated to the memory of Hans Bethe}\\
\end{flushright}
\vspace{0.5cm}

\renewcommand{\thefootnote}{\fnsymbol{footnote}}
\setcounter{footnote}{0}

\begin{center}
{\Large\textbf{\mathversion{bold} Complete Spectrum of Long
Operators\\
in ${\mathcal N}=4$ SYM at One Loop}\par} \vspace{1cm}

\textsc{N.~Beisert$^{a}$, V.A.~Kazakov$^{b,}$\footnote{Membre de
l'Institut Universitaire de France}, K.~Sakai$^b$ and
K.~Zarembo$^{c,}$\footnote{Also at ITEP, Moscow, Russia}} \vspace{5mm}

\textit{$^{a}$ Joseph Henry Laboratories, Princeton University,\\
Princeton, NJ 08544, USA} \vspace{3mm}

\textit{$^{b}$
Laboratoire de Physique Th\'eorique\\
de l'Ecole Normale Sup\'erieure et l'Universit\'e Paris-VI,\\
Paris, 75231, France}\vspace{3mm}

\textit{$^{c}$
Department of Theoretical Physics,\\
Uppsala University, 751 08 Uppsala, Sweden}
\vspace{5mm}

\texttt{nbeisert@princeton.edu}\\
\texttt{kazakov,sakai@lpt.ens.fr}\\
\texttt{konstantin.zarembo@teorfys.uu.se}\par\vspace{1cm}

\vfill

\textbf{Abstract}\vspace{5mm}

\begin{minipage}{12.7cm}

We construct the complete spectral curve for an arbitrary local operator,
including fermions and covariant derivatives,
of one-loop $\mathcal{N}=4$ gauge theory in the thermodynamic limit.
This curve perfectly reproduces the Frolov--Tseytlin limit
of the full spectral curve of classical strings on $AdS_5\times S^5$
derived in \texttt{\arxivno{hep-th/0502226}}.
To complete the comparison we introduce stacks, novel bound states
of roots of different flavors which arise in the thermodynamic limit of
the corresponding Bethe ansatz equations.
We furthermore show the equivalence of various types of Bethe equations
for the underlying $\alg{su}(2,2|4)$ superalgebra,
in particular of the type ``Beauty{}'' and ``Beast{}''.

\end{minipage}

\vspace*{\fill}

\end{center}

\newpage
\setcounter{page}{1}
\renewcommand{\thefootnote}{\arabic{footnote}}
\setcounter{footnote}{0}


\section{Introduction}
\label{sec:Intro}

Non-abelian gauge theories discovered more than 40 years ago still
do not have a satisfactory quantitative description at the intermediate
and strong coupling regimes, in spite of their great importance for
fundamental physics and substantial efforts of many theoretical and
computational physicists and mathematicians. The most important of
them, QCD and the standard model, are well studied perturbatively and
to some extent we know the qualitative picture of strong coupling
phenomena, such as confinement or instanton-induced processes.
Nowadays lattice calculations provide some reasonable quantitative data
confirming the qualitative picture, but they cannot replace systematic
analytical methods, still absent.

A new boost to the study of these questions was given by supersymmetry:
Supersymmetric Yang-Mills (SYM) theories appeared to have special BPS sectors
where certain physical quantities,
protected from renormalization by supersymmetry, can be computed exactly.
Outstanding examples are the Seiberg--Witten low-energy effective action
in $\superN=2$ SYM and the Dijkgraaf--Vafa effective potential
in $\superN=1$ SYM.
This shed a good deal of light on the role of confinement and of the
Higgs mechanism in strongly coupled supersymmetric gauge theories.
Still, the BPS sector is only a tiny fraction of
the theory, the rest being as difficult to access
as in non-supersymmetric theories.

New hope came from an issue which was not expected to play any role
in interacting four-dimensional theories: Quantum integrability,
discovered in a pioneering work of Hans Bethe \cite{Bethe:1931hc} in
relation to the Heisenberg XXX chain, usually applicable exclusively
to exact solutions in (1+1)-dimensional theories, made its
breakthrough into four-dimensional large-$N\indup{c}$ gauge
theories. Here, the simplifications of large-$N\indup{c}$ lead to an
essentially two-dimensional description of parts of the theory
bypassing a no-go theorem for four dimensions. The first traces of
integrability were observed and proved by Lipatov for reggeized
gluons in QCD \cite{Lipatov:1993qn,Lipatov:1993yb}. In
\cite{Lipatov:1994xy} the equivalence of the reggeon Hamiltonian and
the Hamiltonian of the Heisenberg XXX$_0$ spin chain was shown. This
result was independently derived by Faddeev and Korchemsky in
\cite{Faddeev:1994zg} where also the Bethe and Baxter equations for
this model are analyzed.

The observations of integrability look especially interesting and
hopeful in the maximally supersymmetric gauge theory, $\superN=4$
SYM. Again it was Lipatov who first noticed that the evolution
equation for quasi-partonic operators (introduced in
\cite{Bukhvostov:1985rn}) at one loop and in the large-$N\indup{c}$
limit is equivalent to the Hamiltonian of a Heisenberg XXX spin chain
\cite{Lipatov:1997vu} (see also \cite{Lipatov:1998as}).%
\footnote{Appealing to the superconformal nature of $\superN=4$ SYM
we would translate this statement to: The planar one-loop dilatation
operator in the $\alg{sl}(2)$-sector (see \cite{Beisert:2003jj}) is
equivalent to the Hamiltonian of quantum spin chain
with $\alg{sl}(2)$ symmetry.}
Subsequently, integrability of evolution equations for
quasi-partonic operators was found also in QCD and less
supersymmetric theories
\cite{Braun:1998id,Braun:1999te,Belitsky:1999bf} (see
\cite{Belitsky:2004cz} for a recent review). The more recent
achievements gave a boost to the search for integrability in
$\superN=4$ SYM: The planar one-loop matrix of anomalous dimensions
for operators made from the six scalar fields was constructed in
\cite{Minahan:2002ve} and its integrability was shown. The
eigenvalues of the matrix can thus be obtained using the Bethe
ansatz for quantum spin chains with $\alg{so}(6)$ symmetry. The
one-loop dilatation operator for the full theory, i.e.~including
fermions and derivatives, was derived in \cite{Beisert:2003jj}.
Integrability allows to diagonalize the planar dilatation operator
by means of an $\alg{su}(2,2|4)$ algebraic Bethe ansatz
\cite{Beisert:2003yb} and thus obtain all the one-loop anomalous
dimensions for all single-trace operators of any length (defined as
the number of constituent fields).

All these results were at leading non-trivial order in the coupling constant.
Integrability beyond one loop was first observed in \cite{Beisert:2003tq}
and conjectured to hold to all orders.
The restriction to particular closed sectors of the $\superN=4$ SYM theory
allowed to derive the dilatation operator up to five loops
under some reasonable assumptions
\cite{Beisert:2003tq,Beisert:2003jb,Beisert:2004hm,Beisert:2004ry}.
It now seems quite plausible to believe that integrability
persists indeed up to at least three loops in the whole theory
\cite{Beisert:2003ys,Staudacher:2004tk}.
This view is substantiated by some direct calculations
of anomalous dimensions \cite{Kotikov:2003fb,Eden:2004ua}
and an extrapolation \cite{Kotikov:2004er} from an
explicit three-loop calculation in QCD \cite{Vogt:2004mw};
these results agree perfectly with the predictions of integrability!
However, it is hard to imagine that we will discover the whole
integrable structure of $\superN=4$ SYM, if such exists,
by just painfully computing higher and higher orders of
the dilatation operator or of some anomalous
dimensions. The calculations become too cumbersome and virtually
impossible beyond the third order,
although some progress \cite{Klose:2003qc,Fischbacher:2004iu}
was made in the case of the technically quite similar
plane-wave matrix model \cite{Berenstein:2002jq}
which can be thought of as a dimensionally reduced
gauge theory \cite{Kim:2003rz}.

In this situation, a great help can hopefully be provided by the
AdS/CFT correspondence claiming the weak/strong duality of
$\superN=4$ SYM to the IIB string theory on the $AdS_5\times S^5$
background \cite{Maldacena:1998re,Gubser:1998bc,Witten:1998qj}.
Although a weak/strong duality usually prevents quantitative
comparisons in sectors not protected by supersymmetry, two proposals
were made how this problem might be avoided: the plane-wave
correspondence by Berenstein, Maldacena and Nastase
\cite{Berenstein:2002jq} and the spinning strings proposal by Frolov
and Tseytlin \cite{Frolov:2003qc} (see
\cite{Gubser:2002tv,Frolov:2002av} and
\cite{Russo:2002sr,Minahan:2002rc} for earlier qualitative and
quantitative proposals involving particular spinning string
configurations). Both of them involve states with a large spin
quantum number on $S^5$ or of $\alg{so}(6)$ and an effective
coupling constant which can be chosen to be small in both theories.
Although the proposals turn out not to be applicable in a strict
sense --- both theories yield different results starting at third
order in the effective coupling \cite{Callan:2003xr,Serban:2004jf}
--- they have unearthed a striking similarity between gauge theory
and string theory \cite{Beisert:2003xu,Frolov:2003xy,Beisert:2003ea}
(see \cite{Beisert:2004ry,Tseytlin:2003ii,Tseytlin:2004cj,Beisert:2004yq,Zarembo:2004hp} for reviews). The use
of coherent states for gauge theory enabled to make comparisons at
the level of the Hamiltonians and thus show the equivalence of parts
of the spectra \cite{Kruczenski:2003gt,Kruczenski:2004kw,Mikhailov:2004xw}
(and references in \cite{Tseytlin:2004xa}).

It occurs that, just as $\superN=4$ SYM, non-interacting IIB string
theory on $AdS_5\times S^5$ is integrable. At the classical level
this was shown by Bena, Polchinski and Roiban \cite{Bena:2003wd},
but most likely it remains true in the quantum regime (see
\cite{Arutyunov:2004vx,Beisert:2004jw,Swanson:2004qa,Arutyunov:2004yx,
Berkovits:2004xu,Beisert:2005mq,Hernandez:2005nf} for some
indications). Integrability leads to a wide range of applications,
most importantly in this context we can use it as a means of
comparison. The classical integrable structure is closely related to
the one of gauge theory in the thermodynamic limit of long operators
\cite{Arutyunov:2003rg,Kazakov:2004qf,Kazakov:2004nh,Arutyunov:2004xy,Beisert:2004ag,
Beisert:2005bm,Alday:2005gi,Arutyunov:2004yx,Schafer-Nameki:2004ik}
(see also \cite{Arutyunov:2003uj,Dolan:2003uh}).

In this way we see that the AdS/CFT correspondence in combination
with integrability on both sides of the strong/weak duality
strengthens our hope that the $\superN=4$ superconformal gauge
theory is integrable in the whole range of the 't Hooft coupling
$\lambda=N\gym^2$. Let us note that even without the relation to
strings, integrability in gauge theory is of great importance.
Whereas direct or indirect perturbative
computations become virtually impossible after the first few orders,
a Bethe ansatz which applies to arbitrarily high orders might still
have a reasonably simple form \cite{Beisert:2004hm}. Moreover, the
Bethe ansatz is very useful for simplifying the calculation of
anomalous dimensions of operators with many constituent fields,
i.e.~in the thermodynamic limit. {}From a phenomenological point of
view these operators look more exotic, but they are at the heart of
the comparison to Frolov--Tseytlin spinning strings. The main
importance of integrability lies in the possibility of comparing the
structure of gauge theory with that of string theory in the space of
\emph{all operators} and of trying to spread it to all strengths of
the coupling. We hope that this improves our understanding of the
non-protected sector in $\superN=4$ gauge theory.

To advance this program we found it useful to analyze
the thermodynamic limit of one-loop planar $\superN=4$ SYM
for the full space of local operators.
Using the Bethe ansatz equations diagonalizing the matrix of
anomalous dimensions we construct the algebraic curve describing a
complete and rather explicit solution for the dimensions of all long
operators of the theory. Then we compare this curve with the one
obtained in our recent paper \cite{Beisert:2005bm}
and find a perfect coincidence of the two curves in the Frolov--Tseytlin limit.

One important finding of the present paper, which completes
the successful comparison to string theory in \cite{Beisert:2005bm},
is the discovery of apparently new
bound states of Bethe roots in the thermodynamic limit.
These involve different flavors,
i.e.~the Bethe roots belong to different nodes
of the Dynkin diagram of the underlying superalgebra.
We refer to these bound states as \emph{stacks},
see \figref{fig:stack} for an illustration.
An important feature of stacks is that they
can form long supports which we call \emph{strings of stacks}.%
\footnote{The term ``string{}'' here refers to so-called ``Bethe
strings{}'', a set of Bethe roots stretched out in the spectral
parameter plane.} 
They manifest as cuts which connect
non-neighboring sheets of the Riemann surface of the algebraic
curve, similar to the cuts of the finite-gap solution for the string
presented in \cite{Beisert:2004ag,Beisert:2005bm}. In this way we
closed the gap in our understanding of the general structure of the
algebraic curve of gauge theory.%
\footnote{This was still missing in
\cite{Beisert:2004ag}, where we had no evidence for stacks and no
cuts relating remote sheets of the curve.}
Another new phenomenon which we observe in this paper is the \emph{anomaly},
a short distance effect giving in principle a contribution from the
nearby roots.%
\footnote{A similar effect was observed long ago in
matrix models \cite{Matytsin:1993iq} and 2D YM theory
\cite{Daul:1993xz,Boulatov:1993zs}.}
The existence of a simple form
of Bethe equations for stacks is possible due to the cancellation of
this \emph{anomaly}.  Although the anomaly proposed here  cancels at
leading order, its   effects appear to be important in the
corrections to the thermodynamic limit
\cite{Beisert:2005mq,Hernandez:2005nf}.

We clarify the role of fermions in our construction of the algebraic
curve, where they manifest themselves as certain solitary pole
singularities. It is interesting that
in the superstring curve the residues of these poles are nilpotent
as they are bilinears of fermionic coordinates.
This feature permits us to consistently project them out
(if this is our desire).
In gauge theory we can also neglect the fermions,
albeit for a slightly different reason:
In the gauge Bethe equations they appear as
excitations with Fermi-statistics.
They are excluded from forming condensates
and consequently there can only be a few of these,
but not a sufficient amount to contribute to the
leading order of the thermodynamic limit.
Nevertheless, we shall not drop the fermions,
after all they are a very important ingredient of the theory,
but carry them along and show that even including them,
the agreement between gauge and string theory is
perfect.

Another observation of this paper concerns
the non-uniqueness of the Dynkin diagram,
or the set of simple roots,%
\footnote{We hope the reader will not confuse the Bethe roots,
also called rapidities, with the roots of Lie algebra.}
for superalgebras.
We show the equivalence of the algebraic Bethe ans\"atze corresponding
to different bases of superunitary algebras in general
and $\alg{su}(2,2|4)$ in particular.
This allows us to construct solutions for
anomalous dimensions starting from different spin-chain vacua.
In particular, we find a chain of transformations for
Bethe equations from the ``Beauty{}'' to the
``Beast{}'' basis (see \cite{Beisert:2003yb}),
corresponding to the vacua given by the BPS states $\Phi^L$
and non-BPS states $\fldF^L$, respectively.
Another transformation of this type
interchanges the two bosonic factors,
$\alg{su}(4)$ and $\alg{su}(2,2)$, of the superalgebra
while leaving the $\Phi^L$ vacuum unchanged.
These dualities can also provide useful information
for relating different closed sectors of the theory.

The paper is organized as follows: \secref{sec:Gauge} contains a
brief review of the Bethe ansatz for the $\alg{su}(2,2|4)$
superalgebra and its relation to one-loop dimensions of single-trace
operators in $\superN=4$ SYM; in \secref{sec:Dualize} we present the
duality transformations between different types of Bethe equations;
in \secref{sec:Stacks} we consider the thermodynamic limit of Bethe
equations and introduce the stacks as bound states of roots of
different flavors; the Bethe equations in this limit are rewritten
in terms of densities of stacks and the complete factorization of
the bosonic $\alg{su}(4)$ and $\alg{su}(2,2)$ sectors of the theory
is shown; these sectors can interact only through fermions which
appear in corrections to the thermodynamic limit;
in \secref{sec:GaugeCurve} we construct the algebraic
curve for the full one-loop theory in the thermodynamic limit of
long operators extending the work \cite{Schafer-Nameki:2004ik};
in \secref{sec:Compare} we compare this curve to the
algebraic curve in the Frolov--Tseytlin limit of the string theory
and find a perfect coincidence proving another impressive
manifestation of the similarity of string and gauge theories first
noticed by 't Hooft \cite{'tHooft:1974jz}. There are three
appendices: \appref{sec:Beauty} contains lengthy but useful formulae
related to the Bethe ansatz description of the $\alg{su}(2,2|4)$
spin chain; in \appref{sec:Spectral} we describe the characteristic
determinant of the quantum spin chain, in particular from the perspective of
using it to generate transfer matrix eigenvalues;
in \appref{sec:anom} we discuss the anomaly arising from short distance
configurations of Bethe roots in different flavors.

\section{Review of the One-Loop Bethe Ansatz}
\label{sec:Gauge}

In this chapter we shall briefly review one-loop integrability and the
Bethe equations for $\superN=4$ gauge theory \cite{Beisert:2003yb}.
This will set the stage for the sections to come.

\paragraph{States.}

In gauge theory we are interested in
dimensions of single-trace local operators
of the form
\[\label{eq:Gauge.States}
\Op=\Tr \fld{W}_{A_1} \fld{W}_{A_2} \fld{W}_{A_3} \ldots \fld{W}_{A_L}
\]
where $\fld{W}_A$ is any of the following fields of the theory
and their derivatives
\[\label{eq:Gauge.Fields}
\fld{W}_A\in\bigset{\cder^m \Phi,\cder^m \Psi,\cder^m \fld{F}}.
\]
The fields are subject to the equations of motion: Terms like
$\cder\cdot\cder$, $\gamma\cdot\cder$ and $\comm{\cder}{\cder}$ are
not allowed within $\fld{W}_A$ because they can be expressed as some
combination of fields. The number of fields in $\Op$ is called the
length $L$. Within the Bethe ansatz approach the operators correspond
to states of a spin chain. Generically, states will
not be of the simple form suggested in \eqref{eq:Gauge.States},
they will rather be linear combinations of the basis states
presented in \eqref{eq:Gauge.States}.

\paragraph{Energies.}

Local operators in a (super)conformal theory are
characterized by their scaling dimension. Via
the AdS/CFT correspondence, these are dual to energies
of string states. We will therefore refer to scaling dimensions
by the letter $E$.
The dilatation operator $\dil$ measures the scaling dimensions
of local operators
\[
\label{eq:Gauge.DilOp}
\dil\,\Op=E\,\Op.
\]
The one-loop dilatation operator was derived in
\cite{Beisert:2003jj,Beisert:2004ry}.
In the large-$N$ approximation,
this eigenvalue problem can be reformulated
in terms of a quantum spin chain \cite{Minahan:2002ve}
where the dilatation operator represents
the nearest-neighbor spin chain Hamiltonian.
Following \cite{Minahan:2002ve} this spin chain
with $\alg{su}(2,2|4)$ symmetry
was shown to be integrable \cite{Beisert:2003yb}.

\paragraph{Bethe Ansatz.}

\begin{figure}\centering
\includegraphics{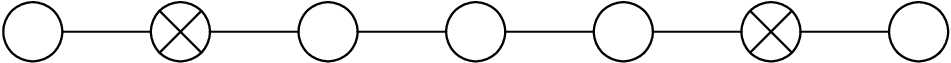}
\caption{``Beauty{}'' \protect\cite{Beisert:2003yb} Dynkin diagram of $\alg{su}(2,2|4)$.}
\label{fig:Beauty}
\end{figure}

Integrability leads to major simplifications in diagonalizing
the Hamiltonian and thus finding the spectrum of scaling dimensions:
The Bethe ansatz provides a set of algebraic equations whose
solutions correspond to eigenstates and their energy alias
scaling dimension can be read off immediately.
An eigenstate within the Bethe ansatz is specified by
a set of Bethe roots $\set{u_p^{(j)}}$.
To formulate the algebraic Bethe ansatz for $\alg{su}(2,2|4)$
we introduce the following seven flavors of Bethe roots
\[
\bigset{u_p^{(j)}}\qquad \mbox{with}\quad j=1,\ldots,7,\quad
p=1,\ldots,K_j
\]
corresponding to 7 nodes of the Dynkin diagram.
Here we make the following choice of grading for the eight
components of the fundamental representation
\[\label{eq:Bethe.Grading}
\sgrad_{3,4,5,6}=+1,
\qquad
\sgrad_{1,2,7,8}=-1.
\]
This means that Bethe roots with $j=2,6$ describe fermionic
excitations while the others represent bosons. This corresponds to
the ``Beauty{}'' choice of the Dynkin diagram of $\alg{su}(2,2|4)$ in \cite{Beisert:2003yb},
c.f.~\figref{fig:Beauty}.

\paragraph{Bethe Equations.}

The Bethe equation for a root $u_p^{(j)}$ reads in a concise form
\[\label{eq:Gauge.Bethe}
\lrbrk{
\frac{u_p^{(j)}+\sfrac{i}{2}V_{j}}
     {u_p^{(j)}-\sfrac{i}{2}V_{j}}}^L
=\prod_{j'=1}^7\prod_{\textstyle\atopfrac{q=1}{(j,p)\neq(j',q)}}^{K_{j'}}
\frac{u_p^{(j)}-u_q^{(j')}+\sfrac{i}{2}M_{j,j'}}
     {u_p^{(j)}-u_q^{(j')}-\sfrac{i}{2}M_{j,j'}}\,,
\]
where $M_{j,j'}$ is the Cartan matrix
corresponding to \figref{fig:Beauty}
and $V_j$ are the Dynkin labels of the
spin representation
\[\label{eq:Gauge.Cartan}
M_{j,j'}=\matr{r|r|rrr|r|r}{
-2&+1&&&&&\\\hline
+1&0&-1&&&&\\\hline
&-1&+2&-1&&&\\
&&-1&+2&-1&&\\
&&&-1&+2&-1&\\\hline
&&&&-1& 0&+1\\\hline
&&&&&+1&-2\\
},\qquad
V_{j}=\matr{c}{
0\\\hline
0\\\hline
0\\
1\\
0\\\hline
0\\\hline
0\\
}.
\]
The equations for all seven types of roots
are spelled out in \appref{sec:Beauty.Gauge}.
They can be viewed as a periodicity condition
on spin excitations propagating around the chain.
In an alternative approach the Bethe equations
are derived as a consistency condition on a
transfer matrix, see \appref{sec:Beauty.Transfer}.
Some transfer matrices can be generated
from the characteristic determinant, c.f.~\appref{sec:Spectral},
which plays a central role in the foundations of
quantum integrable spin chains.

\paragraph{Local Charges.}

States corresponding to single-trace operators of
the $\superN=4$ SYM theory obey, due to cyclicity of the trace,
the zero total momentum condition
\[\label{MOMZ}
\prod_{j=1}^7\prod_{p=1}^{K_j}
\frac{u_p^{(j)}+\sfrac{i}{2}V_j}{u_p^{(j)}-\sfrac{i}{2}V_j}=
\prod_{p=1}^{K_4} \frac{u_p^{(4)}+\sfrac{i}{2}}{u_p^{(4)}-\sfrac{i}{2}}=1.
\]
Note that only the roots of the middle node of the Dynkin diagram
carry momentum because $V_j\neq 0$ only for $j=4$.
The one-loop anomalous dimension $\delta E$
is determined by the (middle) Bethe roots through
\[\label{EDIM}
\delta E=\frac{\lambda}{8\pi^2}\sum_{p=1}^{K_4}
\frac{1}{\bigbrk{u_p^{(4)}}^2+\sfrac{1}{4}}\,.
\]
Finally, the local charges of the integrable spin chain are given by
\[\label{CHARGES}
Q_r=\sum_{p=1}^{K_4}
q_r(u_p^{(4)}),\qquad
q_r(u)
=
\frac{i}{r-1}
\lrbrk{
\frac{1}{(u-\sfrac{i}{2})^{r-1}}-\frac{1}{(u-\sfrac{i}{2})^{r-1}}}
.
\]
Note that the momentum constraint \eqref{MOMZ} and
anomalous dimension \eqref{EDIM} can be written
in terms of the first two charges
as $Q_1=2\pi m$ and $\delta E=(\lambda/8\pi^2)Q_2$.

\paragraph{Global Charges.}

The global charges, i.e.~the eigenvalues of the Cartan
generators of $\alg{su}(2,2|4)$, are determined
by the excitation numbers.
This lengthy relation is presented in \appref{sec:Beauty.Charges}.
The Bethe vacuum with $K_j=0$ represents the state
\[\label{DEFL}
\fldZ^L,
\]
where $\fldZ$ is a complex combination of the scalar fields. The
field $\fldZ$ the highest weight of the irreducible module of all
fields $\fldW_A$ described by the Dynkin labels $V_j$.
For a reasonable highest-weight state the excitation
numbers $K_j$ should satisfy the bounds
\[\label{eq:BOUNDS}
\begin{array}[b]{c}
0\le K_1 \le K_2 \le K_3\le K_4 \ge K_5 \ge K_6 \ge K_7 \ge 0,
\\[0.7ex]
K_2+2L \ge K_3+L\ge K_4 \le K_5+L \le K_6 +2L.
\end{array}
\]
%

\section{Duality Transformation}
\label{sec:Dualize}

Before we engage in the thermodynamic limit and the comparison
to string theory, we shall investigate an important class
of exact duality transformations of the Bethe ansatz.

As opposed to bosonic algebras, the superalgebras allow multiple
inequivalent choices of simple roots related to different orderings
of the gradings within the fundamental representation. This is
reflected by multiple choices of Dynkin diagrams, Cartan matrices
and thus Bethe equations. In this chapter we explain a duality
transformation which connects all the sets of Bethe equations and
therefore show that they are equivalent. In the thermodynamic limit,
discussed in the subsequent sections, this transformation is
trivially realized as a reordering of Riemann sheets, but we can
precisely perform the transformation for the discrete case as well.
Such a transformation is known among statistical physicists as a
particle-hole transformation \cite{Woynarowich:1983aa,Bares:1992aa}
and has been studied for various supersymmetric models solvable by
Bethe ansatz \cite{Essler:1992nk,Essler:1992uc,Gohmannn:2003aa}.
The derivation of the Bethe equations for the $\alg{sl}(2)$ sector
from the ``Beauty'' form of the Bethe ansatz \cite{Staudacher:2004tk}
is a particular case of the general duality transformations (in that
case dual roots do not actually appear).

\subsection{Bethe Equations and Dual Roots}

Suppose that there are $K_u$ Bethe roots $\{u_p\}_{p=1}^{K_u}$
corresponding to a fermionic node of the Dynkin diagram.
Due to the absence of self-interaction terms in the Bethe equations,
we will show how to perform a duality transformation
to eliminate $\{u_p\}$ and rewrite equations in terms
of dual roots $\{\tu_p\}$.
For a generic superunitary algebra, the fermionic roots appear in the
set of Bethe equations as follows, see \cite{Kulish:1985bj}:
\< \label{eq:BE_v} \cdots\cdots\Eqn{=}
\prod_{q=1}^{K_u}\frac{v_p-u_q+\frac{i}{2}}{v_p-u_q-\frac{i}{2}}\cdots\cdots,\\
\label{eq:BE_u} \left(\frac{u_p+\frac{i}{2}V_u}{u_p-\frac{i}{2}V_u}\right)^L \Eqn{=}
\prod_{q=1}^{K_v}\frac{u_p-v_q+\frac{i}{2}}{u_p-v_q-\frac{i}{2}}
\prod_{r=1}^{K_w}\frac{u_p-w_r-\frac{i}{2}}{u_p-w_r+\frac{i}{2}}\,,\\
\label{eq:BE_w} \cdots\cdots\Eqn{=} \prod_{q}^{K_w}
\frac{w_p-u_q-\frac{i}{2}}{w_p-u_q+\frac{i}{2}}\cdots\cdots. \>
Here, the roots $\{v_p\}_{p=1}^{K_v}$
and $\{w_p\}_{p=1}^{K_w}$ correspond to the neighboring nodes
of the Dynkin diagram, i.e.~in the notation of the previous section
$v_p=u^{(j-1)}_p$, $u_p=u^{(j)}_p$, $w_p=u^{(j+1)}_p$.

Let us first consider the case of $V_u\ne 0$.
The set of equations \eqref{eq:BE_u} is equivalent to
the algebraic equation%
\footnote{Strictly speaking, \eqref{eq:BE_u} has $u_p=\infty$ as
a solution while it cannot be seen in \eqref{eq:Pzero}.
Such a root corresponds to descendants
rather than highest weight states, so here we assume no root at $u=\infty$.}
\[\label{eq:Pzero}
P(u_p)=0,
\]
where the polynomial $P(u)$ is given by
\<\label{eq:Pdef} P(u)\eq
\phantom{+}\, (u+\sfrac{i}{2}V_u)^L
\prod_{q=1}^{K_v}(u-v_q-\sfrac{i}{2})
\prod_{r=1}^{K_w}(u-w_r+\sfrac{i}{2})
\nl
-(u-\sfrac{i}{2}V_u)^L
\prod_{q=1}^{K_v}(u-v_q+\sfrac{i}{2})
\prod_{r=1}^{K_w}(u-w_r-\sfrac{i}{2})\\\nonumber
\eq i(V_uL-K_v+K_w)\,u^{L+K_v+K_w-1}+\cdots.
\>
The polynomial $P(u)$ has $L+K_v+K_w-1$ zeros.%
\footnote{It is safe to assume $V_uL-K_v+K_w\ne 0$
for non-trivial and physical states.
Even for $V_uL-K_v+K_w=0$ the dualization can be performed with
a modified number of dualized roots (or alternatively roots
at $u=\infty$).}
By construction
$\{u_p\}_{p=1}^{K_u}$ constitute a part of them, but there are $K_{\tu}$
residual zeros $\{\tu_p\}_{p=1}^{K_\tu}$ with
\[\label{eq:BNTrf}
K_\tu=L+K_v+K_w-K_u-1.
\]
Therefore, $P(u)$ factorizes as
\[\label{eq:Pfact}
P(u)=i(V_uL-K_v+K_w) \prod_{p=1}^{K_u}(u-u_p)\prod_{q=1}^{K_\tu}(u-\tu_q).
\]
Note that each $\tu_p$ satisfies the same Bethe equation
(\ref{eq:BE_u}) as $u_p$. In this sense one can also regard
$\{\tu_p\}_{p=1}^{K_\tu}$ as Bethe roots; they serve as dual
roots to $\{u_p\}_{p=1}^{K_u}$ as we shall see below.

{}From the alternative representations \eqref{eq:Pfact} and \eqref{eq:Pdef} of
the polynomial $P(u)$, it follows that%
\footnote{When $V_u=1$ and there is a root $v_p=0$,
there appear a pair of dual roots $\tu_p=\pm i/2$,
which make the below equations singular.
However, since they always appear together,
one can consistently cancel the divergences
and make the rest finite everywhere.}
\<
\frac{P(v_p+\frac{i}{2})}{P(v_p-\frac{i}{2})}\eq
\prod_{q=1}^{K_u}\frac{v_p-u_q+\frac{i}{2}}{v_p-u_q-\frac{i}{2}}
\prod_{r=1}^{K_\tu}\frac{v_p-\tu_q+\frac{i}{2}}{v_p-\tu_q-\frac{i}{2}}\,,
\nln
\frac{P(v_p+\frac{i}{2})}{P(v_p-\frac{i}{2})}\eq
\left(\frac{v_p-\frac{i}{2}(V_u-1)}{v_p+\frac{i}{2}(V_u-1)}\right)^L
 \prod_{\textstyle\atopfrac{q=1}{q\ne p}}^{K_v}\frac{v_p-v_q+i}{v_p-v_q-i}\,.
\>
Similarly one obtains
\<
\frac{P(w_p-\frac{i}{2})}{P(w_p+\frac{i}{2})}\eq
\prod_{q=1}^{K_u}\frac{w_p-u_q-\frac{i}{2}}{w_p-u_q+\frac{i}{2}}
\prod_{r=1}^{K_\tu}\frac{w_p-\tu_q-\frac{i}{2}}{w_p-\tu_q+\frac{i}{2}}\,,
\nln
\frac{P(w_p-\frac{i}{2})}{P(w_p+\frac{i}{2})}\eq
\left(\frac{w_p-\frac{i}{2}(V_u+1)}{w_p+\frac{i}{2}(V_u+1)}\right)^L
 \prod_{\textstyle\atopfrac{q=1}{q\ne p}}^{K_w}\frac{w_p-w_q-i}{w_p-w_q+i}\,.
\>
By using these equations, one can eliminate $u_p$ from the Bethe
equations \eqref{eq:BE_v,eq:BE_w}
and rewrite them in terms of $\tu_p$.
The same Bethe equation as \eqref{eq:BE_u} holds for $\tu_p$,
we merely invert both sides,
thus the $u_p$-dependent parts of (\ref{eq:BE_v}--\ref{eq:BE_w})
are replaced by%
\footnote{Due to this duality transformation,
there may in principle appear multiple terms on
the l.h.s.~of the Bethe equations.
It does not happen for the spin representation
considered in this work.}
\<
\label{asdflksjf}
 \cdots\cdots\left(\frac{v_p+\frac{i}{2}(V_u-1)}{v_p-\frac{i}{2}(V_u-1)}\right)^L
\Eqn{=} \prod_{\textstyle\atopfrac{q=1}{q\neq p}}^{K_v}\frac{v_p-v_q+i}{v_p-v_q-i}
\prod_{q=1}^{K_\tu}\frac{v_p-\tu_q-\frac{i}{2}}{v_p-\tu_q+\frac{i}{2}}\cdots\cdots,
\nln
\left(\frac{\tu_p-\frac{i}{2}V_u}{\tu_p+\frac{i}{2}V_u}\right)^L
\Eqn{=}
\prod_{q=1}^{K_v}\frac{\tu_p-v_q-\frac{i}{2}}{\tu_p-v_q+\frac{i}{2}}
\prod_{q=1}^{K_w}\frac{\tu_p-w_q+\frac{i}{2}}{\tu_p-w_q-\frac{i}{2}}\,,
\nln
\cdots\cdots\left(\frac{w_p+\frac{i}{2}(V_u+1)}{w_p-\frac{i}{2}(V_u+1)}\right)^L
\Eqn{=}
\prod_{\textstyle\atopfrac{q=1}{q\neq p}}^{K_w} \frac{w_p-w_q-i}{w_p-w_q+i}
\prod_{q=1}^{K_\tu}\frac{w_p-\tu_q+\frac{i}{2}}{w_p-\tu_q-\frac{i}{2}}\cdots\cdots.
\>

Next let us consider the case of $V_u=0$
which leads to different results.
Now the polynomial $P(u)$ is given by
\<\label{eq:Pdef2}
P(u)\Eqn{=}
\prod_{q=1}^{K_v}(u-v_q-\sfrac{i}{2})
\prod_{r=1}^{K_w}(u-w_r+\sfrac{i}{2})
-\prod_{q=1}^{K_v}(u-v_q+\sfrac{i}{2})
\prod_{r=1}^{K_w}(u-w_r-\sfrac{i}{2})
\nln
\Eqn{=} i(-K_v+K_w)\prod_{p=1}^{K_u}(u-u_p)\prod_{q=1}^{K_\tu}(u-\tu_q) \>
and the number of dual roots $\{\tu_p\}$ is
\[\label{eq:BNTrfs0}
K_\tu=K_v+K_w-K_u-1.
\]
The same steps as before lead to the conclusion
that the Bethe equations (\ref{eq:BE_v}--\ref{eq:BE_w})
with $V_u=0$ are replaced by
\<\label{fmvvfdvnf}
 \cdots\cdots\Eqn{=}
\prod_{q=1}^{K_v}\frac{v_p-v_q+i}{v_p-v_q-i}
\prod_{\textstyle\atopfrac{q=1}{q\neq p}}^{K_\tu}\frac{v_p-\tu_q-\frac{i}{2}}{v_p-\tu_q+\frac{i}{2}}\cdots\cdots,
\nln
1\Eqn{=}
\prod_{q=1}^{K_v}\frac{\tu_p-v_q-\frac{i}{2}}{\tu_p-v_q+\frac{i}{2}}
\prod_{q=1}^{K_w}\frac{\tu_p-w_q+\frac{i}{2}}{\tu_p-w_q-\frac{i}{2}}\,,
\nln
\cdots\cdots\Eqn{=}
\prod_{\textstyle\atopfrac{q=1}{q\neq p}}^{K_w} \frac{w_p-w_q-i}{w_p-w_q+i}
\prod_{q=1}^{K_\tu}\frac{w_p-\tu_q+\frac{i}{2}}{w_p-\tu_q-\frac{i}{2}}\cdots\cdots. \>
%

\subsection{Beauty and the Beast}

\begin{figure}[t]
%
%
\begin{minipage}{.49\textwidth}
\begin{center}
\unitlength=2.4pt
\begin{picture}(80,18)
\thicklines \dottedline{.8}( 2.5,10)( 7.5,10)
\dottedline{.8}(12.5,10)(17.5,10) \put(22,10){\line(1,0){6}}
\put(32,10){\line(1,0){6}} \put(42,10){\line(1,0){6}}
\put(52,10){\line(1,0){6}} \dottedline{.8}(62.5,10)(67.5,10)
\dottedline{.8}(72.5,10)(77.5,10) \put(10,10){\circle{4}}
\put(20,10){\circle{4}} \put(30,10){\circle{4}}
\put(40,10){\circle{4}} \put(50,10){\circle{4}}
\put(60,10){\circle{4}} \put(70,10){\circle{4}}
\put(17.69,8.53){\large$\times$} \put(57.69,8.53){\large$\times$}
\put(37,14){$+1$} \put(58.5,3.5){$\dualityarrow$}
\end{picture}\\
\begin{picture}(80,18)
\thicklines \dottedline{.8}( 2.5,10)( 7.5,10)
\dottedline{.8}(12.5,10)(17.5,10) \put(22,10){\line(1,0){6}}
\put(32,10){\line(1,0){6}} \put(42,10){\line(1,0){6}}
\dottedline{.8}(52.5,10)(57.5,10) \put(62,10){\line(1,0){6}}
\dottedline{.8}(72.5,10)(77.5,10) \put(10,10){\circle{4}}
\put(20,10){\circle{4}} \put(30,10){\circle{4}}
\put(40,10){\circle{4}} \put(50,10){\circle{4}}
\put(60,10){\circle{4}} \put(70,10){\circle{4}}
\put(17.69,8.53){\large$\times$} \put(47.69,8.53){\large$\times$}
\put(57.69,8.53){\large$\times$} \put(67.69,8.53){\large$\times$}
\put(37,14){$+1$} \put(48.5,3.5){$\dualityarrow$}
\put(68.5,3.5){$\dualityarrow$}
\end{picture}\\
\begin{picture}(80,18)
\thicklines \dottedline{.8}( 2.5,10)( 7.5,10)
\dottedline{.8}(12.5,10)(17.5,10) \put(22,10){\line(1,0){6}}
\put(32,10){\line(1,0){6}} \dottedline{.8}(42.5,10)(47.5,10)
\put(52,10){\line(1,0){6}} \dottedline{.8}(62.5,10)(67.5,10)
\put(72,10){\line(1,0){6}} \put(10,10){\circle{4}}
\put(20,10){\circle{4}} \put(30,10){\circle{4}}
\put(40,10){\circle{4}} \put(50,10){\circle{4}}
\put(60,10){\circle{4}} \put(70,10){\circle{4}}
\put(17.69,8.53){\large$\times$} \put(37.69,8.53){\large$\times$}
\put(47.69,8.53){\large$\times$} \put(57.69,8.53){\large$\times$}
\put(67.69,8.53){\large$\times$} \put(37,14){$+1$}
\put(38.5,3.5){$\dualityarrow$} \put(58.5,3.5){$\dualityarrow$}
\end{picture}\\
\begin{picture}(80,18)
\thicklines \dottedline{.8}( 2.5,10)( 7.5,10)
\dottedline{.8}(12.5,10)(17.5,10) \put(22,10){\line(1,0){6}}
\dottedline{.8}(32.5,10)(37.5,10) \put(42,10){\line(1,0){6}}
\dottedline{.8}(52.5,10)(57.5,10) \put(62,10){\line(1,0){6}}
\put(72,10){\line(1,0){6}} \put(10,10){\circle{4}}
\put(20,10){\circle{4}} \put(30,10){\circle{4}}
\put(40,10){\circle{4}} \put(50,10){\circle{4}}
\put(60,10){\circle{4}} \put(70,10){\circle{4}}
\put(17.69,8.53){\large$\times$} \put(27.69,8.53){\large$\times$}
\put(37.69,8.53){\large$\times$} \put(47.69,8.53){\large$\times$}
\put(57.69,8.53){\large$\times$} \put(27,14){$+2$} \put(37,14){$-1$}
\put(28.5,3.5){$\dualityarrow$} \put(48.5,3.5){$\dualityarrow$}
\end{picture}\\
\begin{picture}(80,18)
\thicklines \dottedline{.8}( 2.5,10)( 7.5,10)
\dottedline{.8}(12.5,10)(17.5,10) \dottedline{.8}(22.5,10)(27.5,10)
\put(32,10){\line(1,0){6}} \dottedline{.8}(42.5,10)(47.5,10)
\put(52,10){\line(1,0){6}} \put(62,10){\line(1,0){6}}
\put(72,10){\line(1,0){6}} \put(10,10){\circle{4}}
\put(20,10){\circle{4}} \put(30,10){\circle{4}}
\put(40,10){\circle{4}} \put(50,10){\circle{4}}
\put(60,10){\circle{4}} \put(70,10){\circle{4}}
\put(27.69,8.53){\large$\times$} \put(37.69,8.53){\large$\times$}
\put(47.69,8.53){\large$\times$} \put(17,14){$+3$} \put(27,14){$-2$}
\put(38.5,3.5){$\dualityarrow$}
\end{picture}\\
\begin{picture}(80,18)
\thicklines \dottedline{.8}( 2.5,10)( 7.5,10)
\dottedline{.8}(12.5,10)(17.5,10) \dottedline{.8}(22.5,10)(27.5,10)
\dottedline{.8}(32.5,10)(37.5,10) \put(42,10){\line(1,0){6}}
\put(52,10){\line(1,0){6}} \put(62,10){\line(1,0){6}}
\put(72,10){\line(1,0){6}} \put(10,10){\circle{4}}
\put(20,10){\circle{4}} \put(30,10){\circle{4}}
\put(40,10){\circle{4}} \put(50,10){\circle{4}}
\put(60,10){\circle{4}} \put(70,10){\circle{4}}
\put(37.69,8.53){\large$\times$} \put(17,14){$+3$} \put(27,14){$-2$}
\end{picture}\\
\begin{picture}(80,18)
\thicklines \put( 2,10){\line(1,0){6}} \put(12,10){\line(1,0){6}}
\put(22,10){\line(1,0){6}} \put(32,10){\line(1,0){6}}
\dottedline{.8}(42.5,10)(47.5,10) \dottedline{.8}(52.5,10)(57.5,10)
\dottedline{.8}(62.5,10)(67.5,10) \dottedline{.8}(72.5,10)(77.5,10)
\put(10,10){\circle{4}} \put(20,10){\circle{4}}
\put(30,10){\circle{4}} \put(40,10){\circle{4}}
\put(50,10){\circle{4}} \put(60,10){\circle{4}}
\put(70,10){\circle{4}} \put(37.69,8.53){\large$\times$}
\put(17,14){$-3$} \put(27,14){$+2$}
\end{picture}
\caption{From Beauty to Beast\label{fig:B2B}}
\end{center}
\end{minipage}
%
%
\begin{minipage}{.49\textwidth}
\begin{center}
\unitlength=2.4pt
\begin{picture}(80,18)
\end{picture}\\
\begin{picture}(80,18)
\end{picture}\\
\begin{picture}(80,18)
\thicklines \dottedline{.8}( 2.5,10)( 7.5,10)
\dottedline{.8}(12.5,10)(17.5,10) \put(22,10){\line(1,0){6}}
\put(32,10){\line(1,0){6}} \put(42,10){\line(1,0){6}}
\put(52,10){\line(1,0){6}} \dottedline{.8}(62.5,10)(67.5,10)
\dottedline{.8}(72.5,10)(77.5,10) \put(10,10){\circle{4}}
\put(20,10){\circle{4}} \put(30,10){\circle{4}}
\put(40,10){\circle{4}} \put(50,10){\circle{4}}
\put(60,10){\circle{4}} \put(70,10){\circle{4}}
\put(17.69,8.53){\large$\times$} \put(57.69,8.53){\large$\times$}
\put(37,14){$+1$} \put(18.5,3.5){$\dualityarrow$}
\put(58.5,3.5){$\dualityarrow$}
\end{picture}\\
\begin{picture}(80,18)
\thicklines \dottedline{.8}( 2.5,10)( 7.5,10)
\put(12,10){\line(1,0){6}} \dottedline{.8}(22.5,10)(27.5,10)
\put(32,10){\line(1,0){6}} \put(42,10){\line(1,0){6}}
\dottedline{.8}(52.5,10)(57.5,10) \put(62,10){\line(1,0){6}}
\dottedline{.8}(72.5,10)(77.5,10) \put(10,10){\circle{4}}
\put(20,10){\circle{4}} \put(30,10){\circle{4}}
\put(40,10){\circle{4}} \put(50,10){\circle{4}}
\put(60,10){\circle{4}} \put(70,10){\circle{4}} \put(
7.69,8.53){\large$\times$} \put(17.69,8.53){\large$\times$}
\put(27.69,8.53){\large$\times$} \put(47.69,8.53){\large$\times$}
\put(57.69,8.53){\large$\times$} \put(67.69,8.53){\large$\times$}
\put(37,14){$+1$} \put( 8.5,3.5){$\dualityarrow$}
\put(28.5,3.5){$\dualityarrow$} \put(48.5,3.5){$\dualityarrow$}
\put(68.5,3.5){$\dualityarrow$}
\end{picture}\\
\begin{picture}(80,18)
\thicklines \put( 2,10){\line(1,0){6}}
\dottedline{.8}(12.5,10)(17.5,10) \put(22,10){\line(1,0){6}}
\dottedline{.8}(32.5,10)(37.5,10) \dottedline{.8}(42.5,10)(47.5,10)
\put(52,10){\line(1,0){6}} \dottedline{.8}(62.5,10)(67.5,10)
\put(72,10){\line(1,0){6}} \put(10,10){\circle{4}}
\put(20,10){\circle{4}} \put(30,10){\circle{4}}
\put(40,10){\circle{4}} \put(50,10){\circle{4}}
\put(60,10){\circle{4}} \put(70,10){\circle{4}} \put(
7.69,8.53){\large$\times$} \put(17.69,8.53){\large$\times$}
\put(27.69,8.53){\large$\times$} \put(47.69,8.53){\large$\times$}
\put(57.69,8.53){\large$\times$} \put(67.69,8.53){\large$\times$}
\put(37,14){$+1$} \put(18.5,3.5){$\dualityarrow$}
\put(58.5,3.5){$\dualityarrow$}
\end{picture}\\
\begin{picture}(80,18)
\thicklines \put( 2,10){\line(1,0){6}} \put(12,10){\line(1,0){6}}
\dottedline{.8}(22.5,10)(27.5,10) \dottedline{.8}(32.5,10)(37.5,10)
\dottedline{.8}(42.5,10)(47.5,10) \dottedline{.8}(52.5,10)(57.5,10)
\put(62,10){\line(1,0){6}} \put(72,10){\line(1,0){6}}
\put(10,10){\circle{4}} \put(20,10){\circle{4}}
\put(30,10){\circle{4}} \put(40,10){\circle{4}}
\put(50,10){\circle{4}} \put(60,10){\circle{4}}
\put(70,10){\circle{4}} \put(17.69,8.53){\large$\times$}
\put(57.69,8.53){\large$\times$} \put(37,14){$+1$}
\end{picture}\\
\begin{picture}(80,18)
\thicklines \dottedline{.8}( 2.5,10)( 7.5,10)
\dottedline{.8}(12.5,10)(17.5,10) \put(22,10){\line(1,0){6}}
\put(32,10){\line(1,0){6}} \put(42,10){\line(1,0){6}}
\put(52,10){\line(1,0){6}} \dottedline{.8}(62.5,10)(67.5,10)
\dottedline{.8}(72.5,10)(77.5,10) \put(10,10){\circle{4}}
\put(20,10){\circle{4}} \put(30,10){\circle{4}}
\put(40,10){\circle{4}} \put(50,10){\circle{4}}
\put(60,10){\circle{4}} \put(70,10){\circle{4}}
\put(17.69,8.53){\large$\times$} \put(57.69,8.53){\large$\times$}
\put(37,14){$-1$}
\end{picture}
\caption{Interchange of $\alg{su}(4)$ and
$\alg{su}(2,2)$\label{fig:su2sl}}
\end{center}
\end{minipage}
\end{figure}

The Bethe equations for the $\alg{su}(2,2|4)$ algebra can be written in
various ways since different Cartan matrices or Dynkin diagrams
define the same superalgebra. The duality transformation connects
all the possible Bethe equations corresponding to different Dynkin
diagrams. For example, one can transform the Bethe equations from the
``Beauty{}'' to the ``Beast{}'' form, c.f.~\cite{Beisert:2003yb}.
The dualization can be described in a compact
schematic notation with the use of a sequence of Dynkin diagrams
as illustrated in \figref{fig:B2B}.
Another example is given in \figref{fig:su2sl},
which shows a way of interchanging $\alg{su}(4)$ and
$\alg{su}(2,2)$ in the Beauty form.

In the figures, we introduce a slightly extended convention for
the various Dynkin diagrams of $\alg{su}(n|m)$.
We use two kinds of lines: a solid line and a dotted one,
which respectively represent $-1$ and $+1$ as an element
of the symmetrized Cartan matrix.
A bosonic node grows two lines of the same kind
while a fermionic node grows two different kinds.
Also we extend the diagram by adding an exterior line on each end.
Now the Dynkin diagram of $\alg{su}(n|m)$
must have $n$ solid lines and $m$ dotted ones (or vice versa).
This notation in fact manifests the order of gradings
in the fundamental representation.
The number above each node is the corresponding element of the
spin representation vector, i.e.~the Dynkin index, and we discard zeros.
Down arrows indicate the nodes on which the Bethe roots are
being dualized.
Note that the final steps in each of \figref{fig:B2B,fig:su2sl} are not
duality transformations but merely the simultaneous inversion of all
Bethe equations.

Duality transformations can be performed on fermionic roots only.
Each transformation interchanges the two lines
attached to the corresponding fermionic node,
in agreement with \eqref{asdflksjf,fmvvfdvnf}.
Consequently, the gradings of the two neighboring nodes
are inverted. If the node carries a non-zero Dynkin index,
the transformation also modifies the Dynkin indices
according to \eqref{asdflksjf}.

\subsection{Local Charges}
\label{sec:duallocal}

Here we study the transformation of local charges,
in particular the momentum and energy,
under dualization.

\paragraph{Momentum.}

The momentum of a single Bethe root
is given by the function
\[
q_1(u,V_u)=\frac{1}{i}\ln\frac{u+\frac{i}{2}V_u}{u-\frac{i}{2}V_u}\,.
\]
Note that only the Bethe roots with non-zero $V_u$
carry non-zero charges, including momentum.
For a single step of the dualization,
it follows from \eqref{eq:Pdef} and \eqref{eq:Pfact} that
\<
\frac{P(+\frac{i}{2}V_{u})}{P(-\frac{i}{2}V_{u})}\Eqn{=}(-1)^{L+1}
\exp\left(i\sum_{p=1}^{K_v}q_1(v_p,V_{u}-1)
         +i\sum_{p=1}^{K_w}q_1(w_p,V_{u}+1)\right),
\nln
\frac{P(+\frac{i}{2}V_{u})}{P(-\frac{i}{2}V_{u})}\Eqn{=}
\exp\left(i\sum_{p=1}^{K_u}q_1(u_p,V_{u})
         +i\sum_{p=1}^{K_\tu}q_1(\tu_p,V_{u})\right). \>
These equations tell us that the total momentum phase
\[
\exp(iQ_1)=
(-1)^{a(L+1)}
\exp\left(i\sum_{j=1}^7\sum_{p=1}^{K_j}q_1(u^{(j)}_p,V_j)\right)
\]
is conserved throughout the duality transformation.
The constant $a$ counts the number of dualizations of
nodes with a non-zero Dynkin label.
For every such dualization,
the exchange statistics of the spin vacuum changes.
This is reflected by the sign $(-1)^{a(L+1)}$
which is required for the correct cyclicity condition
$\exp(iQ_1)=1$.
For the inversion of the Bethe equations in the final steps
in \figref{fig:B2B,fig:su2sl},
we have to flip the signs within the exponent.
This has no effect on the zero-momentum sector.

\paragraph{Energy.}

The energy of a single Bethe root
is expressed through the function
\[
q_2(u,V_u) =\frac{i}{u+\frac{i}{2}V_u}-\frac{i}{u-\frac{i}{2}V_u}\,.
\]
It follows from \eqref{eq:Pdef} and \eqref{eq:Pfact} that
\<
i\,\frac{P'(+\frac{i}{2}V_{u})}{P(+\frac{i}{2}V_{u})}
-i\,\frac{P'(-\frac{i}{2}V_{u})}{P(-\frac{i}{2}V_{u})}\Eqn{=}
\frac{2L}{V_{u}}+\sum_{p=1}^{K_v}q_2(v_p,V_{u}-1)+\sum_{p=1}^{K_w}q_2(w_p,V_{u}+1),
\nln
i\,\frac{P'(+\frac{i}{2}V_{u})}{P(+\frac{i}{2}V_{u})}
-i\,\frac{P'(-\frac{i}{2}V_{u})}{P(-\frac{i}{2}V_{u})}\Eqn{=}
\sum_{p=1}^{K_u}q_2(u_p,V_{u})+\sum_{p=1}^{K_\tu}q_2(\tu_p,V_{u}).
\>
Using this relation one can transform the total energy
in terms of the Beauty roots to that in the Beast ones as
\[
Q_2=\sum_{p=1}^{K_4}q_2(u^{(4)}_p,+1)
=3L-\sum_{p=1}^{\tK_2}q_2(\tu^{(2)}_p,-3)
         -\sum_{p=1}^{\tK_3}q_2(\tu^{(3)}_p,+2)
\]
where $\tK_j$ are related to $K_j$ by \eqref{ENumBy2Bt}.
The minus signs in front of $q_2$ are due to the inversion
of the Bethe equations in the last step.
The final expression reproduces the energy formula
in terms of the Beast roots found in
\cite{Beisert:2003yb}.

\paragraph{Local Charges.}

The generic expressions for the local charges
\[
Q_r=c_r L\pm \sum_{j=1}^{7}\sum_{p=1}^{K_j}q_r(u^{(j)}_{p},V_j)
\]
do not change under dualization either. Only the inversion in the
last step of \figref{fig:B2B,fig:su2sl} flips the sign.
The transformation however requires the constants $c_r$,
corresponding to the charge density of
the vacuum, to be adjusted when the vacuum changes, see above.

\subsection{The $\alg{su}(1,1|2)$ Subsector}

\begin{figure}[t]
%
%
\begin{center}
\unitlength=2.4pt
\begin{picture}(40,18)
\thicklines
\dottedline{.8}( 2.5,10)( 7.5,10)
\put(12,10){\line(1,0){6}}
\put(22,10){\line(1,0){6}}
\dottedline{.8}(32.5,10)(37.5,10)
\put(10,10){\circle{4}}
\put(20,10){\circle{4}}
\put(30,10){\circle{4}}
\put( 7.69,8.53){\large$\times$}
\put(27.69,8.53){\large$\times$}
\put(17,14){$+1$}
\put(28.5,3.5){$\dualityarrow$}
\end{picture}
\phantom{=}
\begin{picture}(40,18)
\end{picture}\\
\begin{picture}(40,18)
\thicklines
\dottedline{.8}( 2.5,10)( 7.5,10)
\put(12,10){\line(1,0){6}}
\dottedline{.8}(22.5,10)(27.5,10)
\put(32,10){\line(1,0){6}}
\put(10,10){\circle{4}}
\put(20,10){\circle{4}}
\put(30,10){\circle{4}}
\put( 7.69,8.53){\large$\times$}
\put(17.69,8.53){\large$\times$}
\put(27.69,8.53){\large$\times$}
\put(17,14){$+1$}
\put( 8.5,3.5){$\dualityarrow$}
\end{picture}
\raisebox{4.05ex}{$=$}
\begin{picture}(40,18)
\thicklines
\dottedline{.8}( 2.5,10)( 7.5,10)
\put(12,10){\line(1,0){6}}
\dottedline{.8}(22.5,10)(27.5,10)
\put(32,10){\line(1,0){6}}
\put(10,10){\circle{4}}
\put(20,10){\circle{4}}
\put(30,10){\circle{4}}
\put( 7.69,8.53){\large$\times$}
\put(17.69,8.53){\large$\times$}
\put(27.69,8.53){\large$\times$}
\put(17,14){$+1$}
\put(18.5,3.5){$\dualityarrow$}
\end{picture}\\
\begin{picture}(40,18)
\thicklines
\put( 2,10){\line(1,0){6}}
\dottedline{.8}(12.5,10)(17.5,10)
\dottedline{.8}(22.5,10)(27.5,10)
\put(32,10){\line(1,0){6}}
\put(10,10){\circle{4}}
\put(20,10){\circle{4}}
\put(30,10){\circle{4}}
\put( 7.69,8.53){\large$\times$}
\put(27.69,8.53){\large$\times$}
\put(17,14){$+1$}
\end{picture}
\phantom{=}
\begin{picture}(40,18)
\thicklines
\dottedline{.8}( 2.5,10)( 7.5,10)
\dottedline{.8}(12.5,10)(17.5,10)
\put(22,10){\line(1,0){6}}
\put(32,10){\line(1,0){6}}
\put(10,10){\circle{4}}
\put(20,10){\circle{4}}
\put(30,10){\circle{4}}
\put(17.69,8.53){\large$\times$}
\put( 7,14){$+2$}
\put(17,14){$-1$}
\end{picture}
\caption{Duality transformation
for $\alg{su}(1,1|2)$\label{fig:su112}}
\end{center}
\end{figure}

Let us present the Bethe equations for the $\alg{su}(1,1|2)$ subsector
(see e.g.~\cite{Beisert:2004ry} for an introduction to subsectors)
as a sample application of the duality transformations.
This subsector is realized in the Beauty basis
by setting the numbers of Bethe roots to
\[
(K_1,K_2,K_3,K_4,K_5,K_6,K_7)=
(0,K_v-1,K_v,K_u,K_w,K_w-1,0).
\]
After duality transformations on the second and the sixth nodes,
c.f.~\cite{Staudacher:2004tk},
no dual Bethe roots emanate in this particular setup
and hence we are left with the following set of Bethe equations:
\<\label{b123}
1\Eqn{=} \prod_{q=1}^{K_u}\frac{v_p-u_q-\frac{i}{2}}{v_p-u_q+\frac{i}{2}}\,,\nln
\left(\frac{u_p+\frac{i}{2}}{u_p-\frac{i}{2}}\right)^L \Eqn{=}
\prod_{q=1}^{K_v}
\frac{u_p-v_q-\frac{i}{2}}{u_p-v_q+\frac{i}{2}}
\prod_{\textstyle\atopfrac{q=1}{q\ne p}}^{K_u}\frac{u_p-u_q+i}{u_p-u_q-i}
\prod_{q=1}^{K_w}\frac{u_p-w_q-\frac{i}{2}}{u_p-w_q+\frac{i}{2}}\,,\nln
1\Eqn{=}
\prod_{q=1}^{K_u}\frac{w_p-u_q-\frac{i}{2}}{w_p-u_q+\frac{i}{2}}\,,
\>
where $v_p=u^{(3)}_p,\ u_p=u^{(4)}_p,\ w_p=u^{(5)}_p$.
Corresponding Dynkin diagram is shown on the top
of Fig.~\ref{fig:su112}.
This expression can be called the $\alg{su}(2)$-favored form
in the sense that it reduces to the $\alg{su}(2)$ Bethe equation
by setting $K_v=K_w=0$.%
\footnote{In this case a prior dualization is unnecessary.}
Further applying a duality transformation
to the roots $w_p$, one obtains
\<\label{dkjfhasd}
1\Eqn{=} \prod_{q=1}^{K_u}\frac{v_p-u_q-\frac{i}{2}}{v_p-u_q+\frac{i}{2}}\,,
\nln
\left(\frac{u_p+\frac{i}{2}}{u_p-\frac{i}{2}}\right)^L \Eqn{=}
\prod_{q=1}^{K_v}\frac{u_p-v_q-\frac{i}{2}}{u_p-v_q+\frac{i}{2}}
\prod_{q=1}^{K_{\tw}}\frac{u_p-\tw_q+\frac{i}{2}}{u_p-\tw_q-\frac{i}{2}}\,,
\nln
1\Eqn{=}
\prod_{q=1}^{K_u}\frac{\tw_p-u_q+\frac{i}{2}}{\tw_p-u_q-\frac{i}{2}}\,.
\>
These equations are in the $\alg{u}(1|1)$-favored form
(\figref{fig:su112}, the middle row).
Applying a duality transformation
to the roots $v_p$ leads to the $\alg{sl}(2)$-favored form
(\figref{fig:su112}, the lower left-hand).
\< 1\Eqn{=}
\prod_{q=1}^{K_u}\frac{\tv_p-u_q+\frac{i}{2}}{\tv_p-u_q-\frac{i}{2}}\,,
\nln
\left(\frac{u_p+\frac{i}{2}}{u_p-\frac{i}{2}}\right)^L \Eqn{=}
\prod_{q=1}^{K_{\tv}}\frac{u_p-\tv_q+\frac{i}{2}}{u_p-\tv_q-\frac{i}{2}}
\prod_{\textstyle\atopfrac{q=1}{q\ne p}}^{K_u}\frac{u_p-u_q-i}{u_p-u_q+i}
\prod_{q=1}^{K_{\tw}}\frac{u_p-\tw_q+\frac{i}{2}}{u_p-\tw_q-\frac{i}{2}}\,,
\nln
1\Eqn{=}
\prod_{q=1}^{K_u}\frac{\tw_p-u_q+\frac{i}{2}}{\tw_p-u_q-\frac{i}{2}}\,.
\>
Finally, we can also dualize the middle node in \eqref{dkjfhasd}
instead of the first.
This leads to the equations
\<\label{eq:su112disting}
\left(\frac{v_p+i}{v_p-i}\right)^L
\Eqn{=}
\prod_{\textstyle\atopfrac{q=1}{q\neq p}}^{K_v}\frac{v_p-v_q-i}{v_p-v_q+i}
\prod_{q=1}^{K_\tu}\frac{v_p-\tu_q+\frac{i}{2}}{v_p-\tu_q-\frac{i}{2}}\,,
\nln
\left(\frac{\tu_p-\frac{i}{2}}{\tu_p+\frac{i}{2}}\right)^L \Eqn{=}
\prod_{q=1}^{K_v}\frac{\tu_p-v_q+\frac{i}{2}}{\tu_p-v_q-\frac{i}{2}}
\prod_{q=1}^{K_{\tw}}\frac{\tu_p-\tw_q-\frac{i}{2}}{\tu_p-\tw_q+\frac{i}{2}}\,,
\nln
1\Eqn{=}
\prod_{\textstyle\atopfrac{q=1}{q\neq p}}^{K_{\tw}} \frac{\tw_p-\tw_q+i}{\tw_p-\tw_q-i}
\prod_{q=1}^{K_{\tu}}\frac{\tw_p-\tu_q-\frac{i}{2}}{\tw_p-\tu_q+\frac{i}{2}}\,.
\>
which correspond to the distinguished basis for $\alg{su}(1,1|2)$
(\figref{fig:su112}, the lower right-hand).

Using the transformation rules in
\eqref{sec:duallocal} we obtain expressions for
the momentum and energy in the distinguished
basis \eqref{eq:su112disting}.
The momentum for the first three cases and the last case is given by
\<
\exp(iQ_1)\eq
\exp\left(i\sum_{p=1}^{K_u}q_1(u_p,+1)\right)
\nln\eq
(-1)^{a(L+1)}
\exp\left(
i\sum_{p=1}^{K_{\tu}}q_1(\tu_p,-1)
+i\sum_{p=1}^{K_{\tv}}q_1(\tv_p,+2)
\right).
\>
The sign is due to the fermionic spin vacuum.
Similarly, the energy reads
\[
Q_2=\sum_{p=1}^{K_u}q_2(u_p,+1)
=2L+\sum_{p=1}^{K_{\tu}}q_2(\tu_p,-1)
         +\sum_{p=1}^{K_{\tv}}q_2(\tv_p,+2).
\]
The fermionic spins contribute a vacuum energy of $2L$.

\section{Stacks and the Thermodynamic Limit}
\label{sec:Stacks}

In this section we shall investigate the thermodynamic limit of the
Bethe equations appropriate for the comparison with string theory.
This scaling limit is specified by long spin chains $L\gg 1$
(i.e.~SYM operators consisting of many fields) with
macroscopically many excitations $K_j=\order{L}$ and in the
low-energy regime $\delta E=\order{1/L}$. In this limit, the roots
scale with $L$ as $u^{(j)}_p=\order{L}$. As we shall see, the roots
are organized in two types of structures: Strings and stacks. A
\emph{string} is a collection of roots $u^{(j)}_p$ of fixed flavor
$j$ which are stretched along a curve in the complex plane. A
\emph{stack} is a collection of roots $u^{(j)}_p$ with consecutive
flavors $j,j+1,j+2,\ldots$ which are all centered around some point in
the complex plane.%
\footnote{A different type of ``stack{}''
was used in \cite{Schafer-Nameki:2004ik}
to explain the thermodynamic limit of
fermionic Bethe roots.
{}From the arguments that follow, this stack configuration
does not seem natural, but we cannot exclude it either.}
In both cases, the typical separation of roots is
$\order{1}$. Furthermore, strings and stacks can be combined into
strings of stacks.

The interpretation of these two types of structures is as follows:
A string corresponds to a (Bose-Einstein)
condensate of excitations. In a stringy picture,
the excitations are string oscillators of a fixed
mode number. Their coherent behavior describes the
macroscopic motion of the string.
The type of stack specifies the orientation
of the particular string oscillators in target space.
There are only a handful of stack types which can be neatly
associated to excitation modes of strings and,
in particular, to the BMN spectrum.

\subsection{Formation of Strings and Stacks}
\label{sec:Stacks.Formation}

To explain qualitatively the possibility of stack formation, we
recall the electrostatic interpretation of the Bethe equations
\eqref{eq:Gauge.Bethe}: If we take their logarithms and expand, as
usual, with respect to the inverse differences of rapidities, they
can be viewed as equilibrium conditions for Coulomb-like particles
$u^{(j)}_p$ on the complex plane. The sign of the force between two
particles is determined by the Cartan matrix
\eqref{eq:Gauge.Cartan}.
Two alike bosonic particles ($j=j'=1,3,4,5,7$) repel each other%
\footnote{For this interpretation, we should invert the sign of rows
$1,7$ in \eqref{eq:Gauge.Cartan}. In general, it is however more
convenient to work with a symmetric matrix $M_{j,j'}$.} while two
alike fermions ($j=j'=2,6$) exert no forces on each other. Particles
with adjacent flavor indices ($j=j'\pm 1$) attract each other and
all other pairs ($|j-j'|>1$) do not feel each other's presence.
Finally, there is a global potential from the l.h.s.~of
\eqref{eq:Gauge.Bethe} for particles with $V_j\neq 0$, i.e.~$j=4$.
These we shall call \emph{momentum-carrying}, the others are called
\emph{auxiliary}.

Bethe roots live in an effective potential created by the other
roots of a given solution and the global potential. They must reside
at a point where the effective potential equals $2\pi n^{(j)}_p$,
which stems from the branch ambiguity in the logarithmic form of the
Bethe equations. Let us start with a momentum-carrying Bethe root
$u^{(4)}_p$. Their equilibrium position is mainly determined by the
global potential and their mode number $n^{(4)}_p$. If there are
many roots with a coincident mode number they will all be forced
to the same region of the complex plane. They will, however, not
collapse onto the same point but, due to mutual repulsion, form
extended structures in the complex plane, the so-called Bethe
strings \cite{Sutherland:1995aa,Beisert:2003xu}.

The situation for an auxiliary root is different.
There is no global potential so its position is
determined through the effective potential of the
other roots alone.%
\footnote{This potential is bounded on a global scale and auxiliary mode
numbers $n^{(j)}_p$ must therefore be sufficiently small.
Within many solutions they are simply zero.}
As there is attraction between adjacent flavors
of roots, the auxiliary roots will attach to some root with adjacent
flavor. In this way, the roots will form extended structures in the
flavor index, which we call stacks. A $(k,l)$-stack contains one
root of each flavor $k\leq j< l$.

Stacks behave much like individual Bethe roots:
Two stacks will either attract, repel or not feel
each other's presence. A stack is either bosonic or fermionic
depending on the number of fermionic roots it contains.
Two alike bosonic stacks repel each other while
two alike fermionic stacks exert no mutual force. A stack is
also either momentum-carrying or auxiliary depending
on whether it contains a momentum-carrying root.
Momentum-carrying stacks feel the global potential, while
auxiliary ones live in the effective potential of the other
stacks.

Just like the individual roots, stacks can form strings.%
\footnote{We shall refer to individual Bethe roots
as stacks (with only one level) as well.}
We assume that the type of stack does not change within
the string.%
\footnote{It is not inconceivable that stacks of
different types can form a single compound.
It even seems that such compounds should dominate to obtain a correct
counting of states.
However, we need not consider this situation separately:
Compounds can be thought of as composed from several
strings of stacks of uniform type.}
Here we have to distinguish three cases depending
on the grading according to \eqref{eq:Bethe.Grading}:
\begin{bulletlist}
\item
Bosonic stacks of positive grading $\sgrad_k=\sgrad_l=+1$ ($k,l=3,4,5,6$)
are associated to the compact $\alg{su}(4)$-part of the
$\alg{su}(2,2|4)$-algebra.
These form strings which stretch out in the complex plane
in the imaginary direction. This is similar to
the solutions considered in \cite{Minahan:2002ve,Beisert:2004ag}.

\item
Bosonic stacks of negative grading $\sgrad_k=\sgrad_l=-1$ ($k,l=1,2,7,8$)
are associated to the non-compact $\alg{su}(2,2)$-part of the
$\alg{su}(2,2|4)$-algebra.
They form strings stretching along the real axis of the complex plane
\cite{Kazakov:2004nh}.

\item
Fermionic stacks which have mixed grading $\sgrad_k\neq\sgrad_l$
($k=3,4,5,6$ and $l=1,2,7,8$ or vice versa) cannot form strings.
There is no mutual repulsion of fermionic stacks,
so two of them would collapse onto precisely the same point in
the complex plane.
However two Bethe roots of the same flavor must not coincide
and in this way the Pauli principle is realized in the context of the
Bethe ansatz.%
\footnote{There is a curious alternative explanation
of the Pauli principle: Strings associated to $\alg{su}(4)$
stretch in the imaginary direction whereas strings associated
to $\alg{su}(2,2)$ stretch along the real axis.
Fermions belong to both worlds and therefore should stretch
in both directions at the same time. The only way out of this paradox
is a single point.}
Note that the exclusion of coincident roots is valid for
bosonic roots as well, but due to mutual repulsion they evade the
Pauli principle.
Therefore fermionic stacks must occupy well-separated points.
Hence we expect only solitary, well-separated fermionic
stacks which have to occupy different mode numbers.
They must also lie on the real axis \cite{Schafer-Nameki:2004ik}.

\end{bulletlist}
While bosonic stacks in general form strings with a
macroscopic number of roots, there can only be a finite number
of fermionic stacks.
A finite number of excitations cannot influence the macroscopic
Bethe strings in the main order of the thermodynamic limit and
solitary stacks contribute only to $\order{1/L}$ corrections.
However, a meaningful problem would be to find the
positions of such roots in a given bosonic background.

Note that we can have infinitely small Bethe strings (consisting of a few
bosonic stacks) which have a role very similar to fermionic stacks.
In what follows, we shall rewrite the Bethe equations in a way where
we treat all these stacks on equal footing and
finally split up the different kinds of stacks.

\subsection{Stacks of Roots}
\label{sec:Stacks.Stacks}

\begin{figure}\centering
\includegraphics{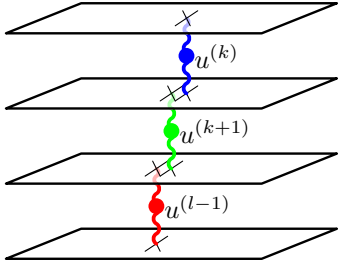}
\caption{A stack of $l-k=3$ Bethe roots $u^{(j)}$, $j=k,\ldots ,l-1$.
The distance between the roots $u^{(j+1)}-u^{(j)}$ is $\order{1}$
while their position $u^{(j)}$ is $\order{L}$.}
\label{fig:stack}
\end{figure}

Now we shall introduce and investigate stacks of Bethe roots in the
thermodynamic limit. A $(k,l)$-stack is a set of Bethe roots
$\set{u^{(j)}}$ of types $k\leq j<l$, which are all close together and
approximated by some $u^{(kl)}$,
c.f.~\figref{fig:stack}. To be more precise,
the center $u^{(kl)}$ is large, $\order{L}$,
while the deviations $u^{(j)}-u^{(kl)}$ within a stack
will be small, $\order{1}$. We shall argue that, for higher-rank
algebras, these are the fundamental objects in the thermodynamic
limit. Furthermore, the precise positions of the roots $u^{(j)}$ are
irrelevant, only the center $u^{(kl)}$ enters the effective Bethe
equations in the thermodynamic limit.

Let us first consider the self-interaction of a stack. A single root
does not self-interact, but there are interactions
$s_{j,j'}(u^{(j)},u^{(j')})$ between the
different constituent roots $u^{(j)}$ of the stack.
Luckily, these have no influence
on the center $u^{(kl)}$ which is again free.
To obtain equations for $u^{(kl)}$
we simply multiply all the Bethe equations for the constituents
$u^{(j)}$. In total we get for the scattering terms
\[
s_{kl,kl}(u^{(kl)},u^{(kl)}):=
\prod_{\textstyle\atopfrac{j,j'=k}{j'\neq j}}^{l-1}
s_{j,j'}(u^{(j)},u^{(j')})=
\prod_{\textstyle\atopfrac{j,j'=k}{j'\neq j}}^{l-1}
\frac{u^{(j)}-u^{{(j')}}+\frac{i}{2}M_{j,j'}}{u^{(j)}-u^{(j')}-\frac{i}{2}M_{j,j'}}=1.
\]
This follows from the antisymmetry of the the scattering phase
of two different constituents of the stack, $s_{j,j'}=s_{j',j}^{-1}$,
due to a symmetric Cartan matrix $M_{j,j'}=M_{j',j}$.

Let us now consider the interaction of two stacks, $\set{u^{(j)}}$ of
type $(k,l)$ and $\set{u^{\prime(j)}}$ of type $(k',l')$.
The combined scattering terms read
\[\label{eq:stackint}
s_{kl,k'l'}(u^{(kl)},u^{\prime(k'l')}):= \prod_{j=k}^{l-1} \prod_{j'=k'}^{l'-1}
\frac{u^{(j)}-u^{\prime(j')}+\frac{i}{2}M_{j,j'}}
{u^{(j)}-u^{\prime(j')}-\frac{i}{2}M_{j,j'}}\,.
\]
When $u^{(kl)}$ and $u^{\prime(k'l')}$ are well-separated,
i.e.~$u^{(kl)}-u^{\prime(k'l')} \sim L$, we can
straight-forwardly compute the thermodynamic limit:
\[\label{eq:asdfl}
s_{kl,k'l'}(u^{(kl)},u^{\prime(k'l')})=
\exp\left(\frac{i M_{kl,k'l'}}{u^{(kl)}-u^{\prime(k'l')}}+ \order{1/L^2}\right),
\qquad M_{kl,k'l'}=\sum_{j=k}^{l-1}\sum_{j'=k'}^{l'-1}M_{j,j'}.
\]
Interestingly, for unitary algebras the double sum is
non-zero only when $\set{k,l}$ and $\set{k',l'}$ have a common element:
Two stacks interact only when they have coinciding labels
\[\label{eq:Stacks.Matrix}
M_{kl,k'l'}= \sgrad_k\lrbrk{\delta_{k,k'}-\delta_{k,l'}}
-\sgrad_l\lrbrk{\delta_{l,k'}-\delta_{l,l'}},
\]
where $\sgrad_k=\pm 1$ describes the grading of the $k$-th component of
the fundamental representation, i.e.~$\sgrad_{1,2,7,8}=-1$ and
$\sgrad_{3,4,5,6}=+1$ \eqref{eq:Bethe.Grading}.

When two stacks are much closer, namely $u^{(kl)}-u^{\prime(k'l')}\sim O(L^0)$,
then strictly speaking \eqref{eq:asdfl} does not apply since we
have to take into account short distance effects
(we will call them anomaly).
This generally happens only within strings of stacks.
A string consists of stacks of the same type,
therefore we should carefully investigate
the case $(k',l')=(k,l)$ when some $u^{(kl)}$ and $u^{\prime(kl)}$ are close,
to see whether we missed some contributions to \eqref{eq:asdfl}.
Fortunately, the exact definition
\eqref{eq:stackint} is manifestly
antisymmetric
\[
s_{kl,k'l'}(u^{(kl)},u^{\prime(kl)})=s^{-1}_{k'l',kl}(u^{\prime(kl)},u^{(kl)}),
\]
in particular $s_{kl,kl}(u^{(kl)},u^{\prime(kl)})=s^{-1}_{kl,kl}(u^{\prime(kl)},u^{(kl)})$.
Therefore there is no anomaly as described in \appref{sec:anom}
when inserting nearby stacks into a string.
We conclude that the approximation \eqref{eq:asdfl} to the formula
\eqref{eq:stackint} is always valid in the thermodynamic limit and
we can study strings of infinitely rigid stacks.

Finally, from the potential terms in the product of Bethe equations
for a stack we obtain in the thermodynamic limit
\[\label{POTEN}
v_{kl}(u^{(kl)}):=
\prod_{j=k}^{l-1}
\lrbrk{\frac{u^{(j)}+\sfrac{i}{2}V_j}{u^{(j)}-\sfrac{i}{2}V_j}}^L=
\exp\lrbrk{\frac{iLV_{kl}}{u^{(kl)}}
+\order{1/L}}\,,
\qquad
V_{kl}=\sum_{j=k}^{l-1}V_j.
\]
Here, $V_{kl}=1$ for the momentum-carrying stacks with
$k=1,2,3,4$ and $l=5,6,7,8$ and
$V_{kl}=0$ otherwise. In analogy to
\eqref{eq:Stacks.Matrix} we can write $V_{kl}$ as
\[\label{eq:Stacks.Vector}
V_{kl}=\half \sheetsign_k-\half \sheetsign_l,\quad
\mbox{with}\quad
\sheetsign_k=(+1,+1,+1,+1,-1,-1,-1,-1).
\]

It is well-understood that Bethe roots are associated to simple
roots of the symmetry algebra. From the above discussion the algebraic
meaning of the stacks becomes clear: They are associated to positive
roots of the algebra.

\subsection{Stacks as Fundamental Excitations}
\label{sec:Stacks.Exitations}

Stacks are labelled by $1\le k<l\le 8$, i.e.~there are 28 types of
stacks. There are 16 fermionic stacks which involve only one of the
two fermionic roots ($\sgrad_k\neq \sgrad_l$).
The remaining 12 bosonic roots can be distinguished by their grading:
6 stacks have positive grading ($\sgrad_k=\sgrad_l=+1$) and
the other 6 have negative grading ($\sgrad_k=\sgrad_l=-1$).
We can split down these numbers even further using
the global potential.
Only stacks with $\sheetsign_k\neq \sheetsign_l$,
i.e.~$k=1,2,3,4$ and $l=5,6,7,8$
can feel the global potential \eqref{POTEN}.
Half of these 16 momentum-carrying stacks are fermionic and
the remaining bosonic stacks split evenly between both gradings.
The momentum-carrying sector thus consists of $4+4$ bosons,
associated to the subalgebras $\alg{su}(4)$ and $\alg{su}(2,2)$,
as well as $8$ fermions.
The remaining 8 fermionic and $2+2$ bosonic modes
do not see the potential ($\sheetsign_k=\sheetsign_l$)
and can be considered auxiliary.

The momentum-carrying stacks agree precisely with the counting of
modes in the BMN limit: A generic BMN operator contains sixteen
types of elementary excitations eight of which are fermionic. A BMN
state is obtained by inserting several of the sixteen types of
excitations into the Bethe vacuum
\[\label{BMNvac}
\state{0}=\mathcal{Z}^L.
\]
To enumerate the excitations a
$\alg{su}(2)\times\alg{su}(2)\times\alg{su}(2)\times\alg{su}(2)$ notation
is very useful:
The six scalars, sixteen fermions and four derivatives
of $\superN=4$ SYM can be written as
\[
\mathcal{Z},\quad
\Phi_{a\dot b},\quad
\bar{\mathcal{Z}},\qquad
\Psi_{\alpha a},\quad
\Psi_{\alpha \dot a},\quad
\Psi_{a\dot\alpha},\quad
\Psi_{\dot a\dot\alpha},\qquad
\mathcal{D}_{\alpha\dot\beta}
\]
where $a,b=1,2$ and $\dot a,\dot b=\dot 1,\dot 2$
are indices of $\alg{su}(2+2)$ and
$\alpha,\beta=\sone,\stwo$ and $\dot \alpha,\dot\beta=\sonedot,\stwodot$
are indices of $\alg{su}(2,2)$.
Often a $\superN=1$ language is employed
where%
\footnote{Conversely one could also write
$\mathcal{Z}=\Phi_{12}$ and
$\bar{\mathcal{Z}}=\Phi_{\dot 1\dot 2}$.}
\[
\Phi_{\dot ab}\sim
\matr{cc}{\bar{\mathcal{X}}&\bar{\mathcal{Y}}\\\mathcal{Y}&\mathcal{X}},
\]
$\Psi_{a\dot\alpha}=\varepsilon_{ac}\bar\Psi^c_{\dot\alpha}$,
$\Psi_{\dot a\dot\alpha}=\varepsilon_{\dot a\dot c}\bar\Psi^{\dot c}_{\dot\alpha}$
as well as $\dot 2=3$, $\dot 1=4$.
The scalar field $\mathcal{Z}$ represents the vacuum.
The elementary excitations are given by the fields
\[
\Phi_{a\dot b},\qquad
\Psi_{\alpha \dot a},\quad
\Psi_{a\dot\alpha},\qquad
\mathcal{D}_{\alpha\dot\beta}\mathcal{Z},
\]
which are to replace one $\mathcal{Z}$ in \eqref{BMNvac}. All other
fields $\bar{\mathcal{Z}}, \Psi_{\alpha a}, \Psi_{\dot\alpha\dot a}$,
a derivative $\mathcal{D}_{\dot\alpha\beta}$ acting on anything
but $\mathcal{Z}$ or multiple derivatives are considered to be
multiple excitations on a single site and thus not elementary. Using a
supersymmetric oscillator $\set{a^\alpha}_{\alpha=\sone}^\stwo$,
$\set{b^{\dot{\alpha}}}_{\dot{\alpha}=\sonedot}^\stwodot$,
$\set{c^a}_{a=1}^2$ and $\set{d^{\dot a}}_{\dot a=\dot 1}^{\dot 2}$
satisfying
\[
\comm{a^\alpha}{a^\dagger_\beta}
=\delta^\alpha_{\beta},\quad
\comm{b^{\dot{\alpha}}}{b^\dagger_{\dot{\beta}}}
=\delta^{\dot{\alpha}}_{\dot{\beta}},\quad
\acomm{c^a}{c^\dagger_b}=\delta^a_{b},\quad
\acomm{d^{\dot a}}{d^\dagger_{\dot b}}=\delta^{\dot a}_{\dot b},
\]
we can also write the elementary excitations as
fields with precisely two oscillator excitations:
\[\label{STATES}
\begin{array}[b]{cccc}
\state{\bar{\mathcal{X}}}=c_1^\dagger d_{\dot 2}^\dagger \state{\mathcal{Z}},&
\state{\mathcal{D}_{\sone\sonedot}\mathcal{Z}}=a_\sone^\dagger b_\sonedot^\dagger\state{\mathcal{Z}},&
\state{\Psi_{\sone 3}}=a_\sone^\dagger d_{\dot 1}^\dagger\state{\mathcal{Z}},&
\state{\bar{\Psi}^{1}_{\sonedot}}=c_2^\dagger b_\sonedot^\dagger \state{\mathcal{Z}},\\[1ex]
\state{\mathcal{Y}}=c_2^\dagger d_{\dot 2}^\dagger\state{\mathcal{Z}},&
\state{\mathcal{D}_{\stwo\sonedot}\mathcal{Z}}=a_\stwo^\dagger b_\sonedot^\dagger\state{\mathcal{Z}},&
\state{\Psi_{\sone 4}}=a_\sone^\dagger d_{\dot 2}^\dagger\state{\mathcal{Z}},&
\state{\bar{\Psi}^2_\sonedot}=c_1^\dagger b_\sonedot^\dagger\state{\mathcal{Z}},\\[1ex]
\state{\mathcal{X}}=c_2^\dagger d_{\dot 1}^\dagger\state{\mathcal{Z}},&
\state{\mathcal{D}_{\sone\stwodot}\mathcal{Z}}=a_\sone^\dagger b_\stwodot^\dagger\state{\mathcal{Z}},&
\state{\Psi_{\stwo 3}}=a_\stwo^\dagger d_{\dot 1}^\dagger\state{\mathcal{Z}},&
\state{\bar{\Psi}^1_\stwodot}=c_2^\dagger b_\stwodot^\dagger\state{\mathcal{Z}},\\[1ex]
\state{\bar{\mathcal{Y}}}=c_1^\dagger d_{\dot 1}^\dagger\state{\mathcal{Z}},&
\state{\mathcal{D}_{\stwo\stwodot}\mathcal{Z}}=a_\stwo^\dagger b_\stwodot^\dagger\state{\mathcal{Z}},&
\state{\Psi_{\stwo 4}}=a_\stwo^\dagger d_{\dot 2}^\dagger\state{\mathcal{Z}},&
\state{\bar{\Psi}^2_\stwodot}=c_1^\dagger b_\stwodot^\dagger\state{\mathcal{Z}}.
\end{array}
\]
The reference state $\state{\mathcal{Z}}$ is annihilated by all
$a^\alpha,b^{\dot{\alpha}},c^{a},d^{\dot a}$.

\begin{figure}\centering
\includegraphics{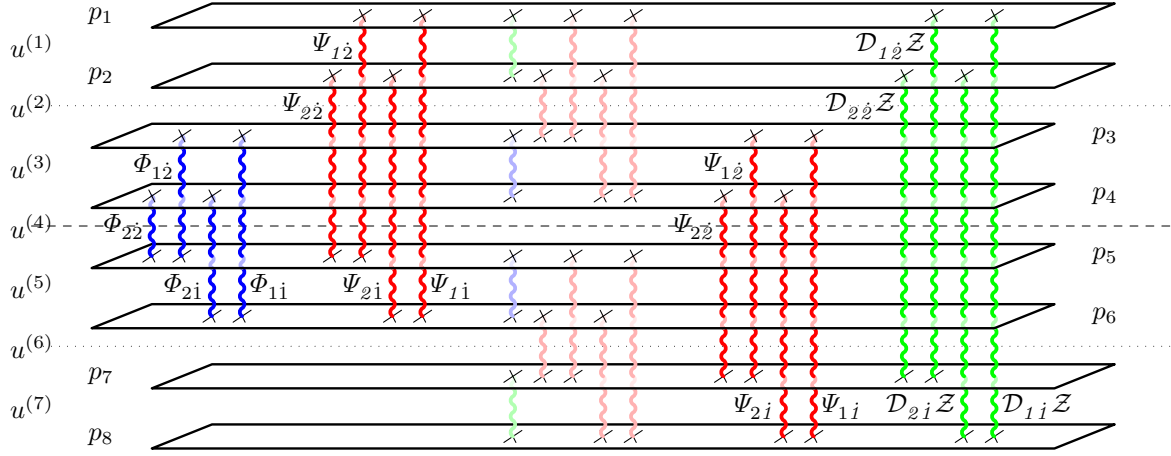}
\caption{
All 28 types of stacks. The outer 16 stacks are momentum-carrying,
the inner 12 are auxiliary.
For all momentum-carrying stacks,
we have indicated the elementary excitation
in terms of fields of $\superN=4$ SYM.
}
\label{fig:dict}
\end{figure}

Let us now relate the momentum carrying stacks to the elementary
excitations, c.f.~\figref{fig:dict}.
The bosonic stacks with $k,l=3,4,5,6$ correspond
to $\alg{su}(4)$ and should thus be related to the scalars
$\Phi_{a\dot b}$.
Analogously, the bosonic stacks with $k,l=1,2,7,8$ correspond
to $\alg{su}(2,2)$ and should be related to the derivatives
$\mathcal{D}_{\alpha\dot \beta}$. Finally, the fermionic
stacks correspond to $\Psi_{\alpha\dot a}$ and $\Psi_{a\dot \alpha}$.
Note that the auxiliary stacks have no direct correspondence as
fields of $\superN=4$ SYM.

\subsection{Strings of Stacks}
\label{sec:Stacks.Compounds}

\begin{figure}\centering
\includegraphics{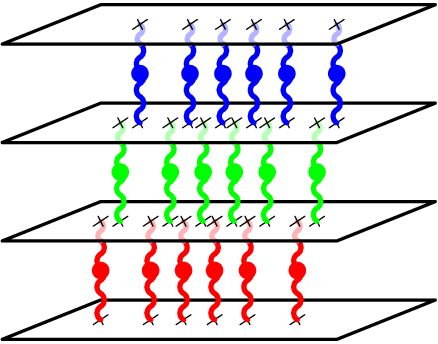}\quad
\includegraphics{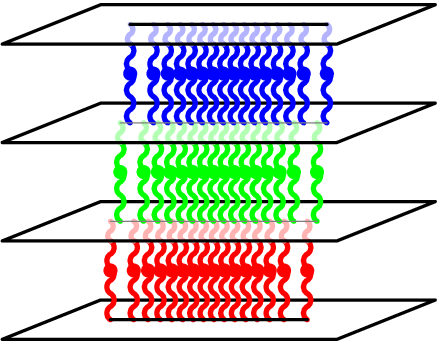}\quad
\includegraphics{bksz.Stacks.Cut.eps}
\caption{A string of stacks as $L$ approaches $\infty$.
{}From left to right: $L$ small, $L$ large, thermodynamic limit.}
\label{fig:string}
\end{figure}

In the scaling limit of long spin chains $L\gg 1$,
corresponding to SYM operators consisting of many fields,
the positions of the stacks scale as $u^{(kl)}_p=\order{L}$.
For each type of stack, introduce a resolvent%
\footnote{We have rescaled the
continuous version $u$ by $L$.}
\[\label{eq:Stacks.ResolvDiscrete}
G_{kl}(u)=\sum_{p=1}^{K_{kl}} \frac{1}{u^{(kl)}_{p}-uL}\,,\qquad
1\le k<l\le 8.
\]
Bosonic stacks may form strings with a density $\rho_{kl}(u)$
along a (disconnected) curve $\contour{C}_{kl}$
as explained in \secref{sec:Stacks.Formation},
c.f.~\figref{fig:string}.
Then the resolvent becomes
\[\label{dfaslkj}
G_{kl}(u)=\int_{\contour{C}_{kl}} \frac{dv\,\rho_{kl}(v)}{v-u}\,.
\]

The Bethe equations are obtained by the usual procedure, but applied
to the stacks, rather than to individual roots: when a stack makes a
tour around the chain its scattering phase after interaction with
all other stacks is $2\pi\Integers$, due to the periodicity:
\[
-\sum_{k'=1}^{7}\sum_{l'=k'+1}^8 M_{kl,k'l'}\resolvsl_{k'l'}(u)
-\frac{V_{kl}}{u} =2\pi n_{kl,a}, \qquad
\mbox{for } u\in \contour{C}_{kl,a}.
\]
When we substitute
\eqref{eq:Stacks.Matrix,eq:Stacks.Vector}
the above equation reads
\[\label{BAEfull}
-\sgrad_{k}\sum_{j=1}^{8}\resolvsl_{kj}(u)
+\sgrad_{l}\sum_{j=1}^{8}\resolvsl_{lj}(u)
-\frac{\sheetsign_k-\sheetsign_l}{2u}=2\pi n_{kl,a},\qquad
\mbox{for }u\in \contour{C}_{kl,a}.
\]
For convenience, we have defined all resolvents as
antisymmetric in the stack indices $(k,l)$: $G_{lk}(u)=-G_{kl}(u)$.
Here $\contour{C}_{kl,a}$ is a connected component of $\contour{C}_{kl}$
and $n_{kl,a}$ is the corresponding mode number. Note that we have
used a slash to indicate a principal value prescription on all resolvents.
In a given equation, it will however only be relevant
for the resolvent $G_{kj}$ with the cut $\contour{C}_{kl,a}$.
The remaining slashed resolvents coincide with
their unslashed value.

Introduce the quasi-momenta
\[\label{QUASIP}
p_k(u)=
\frac{\sheetsign_k}{2u}+ \sgrad_k\sum_{l=1}^{8}G_{kl}(u)
\,,\qquad k=1,\ldots,8.
\]
These functions determine the
eigenvalues of the transfer matrices, see
\appref{sec:Beauty.Transfer} and \eqref{alskfjasd}.
They let us write the Bethe equations as%
\footnote{We have the freedom to add some function $f(u)$
independent of $k$ to all quasi-momenta $p_k(u)$ without changing the
below Bethe equations.}
\[\label{eq:BEAP}
\sheetsl_l(u)-\sheetsl_k(u) =2\pi n_{kl,a}, \qquad \mbox{for } u\in
\contour{C}_{kl,a}.
\]
%

\subsection{Local and Global Charges}
\label{sec:Stacks.Charges}

In the thermodynamic limit we can express the
trace cyclicity \eqref{MOMZ},
the one-loop anomalous dimension \eqref{EDIM}
and the higher integrable charges \eqref{CHARGES}
in terms of the resolvents \eqref{eq:Stacks.ResolvDiscrete}
\[\label{ZeroQs}
G\indup{mom}(0)=2\pi m,
\qquad
\delta E=\frac{\lambda}{8\pi^2 L}\, G\indup{mom}'(0).
\qquad
G\indup{mom}(u)=\sum_{r=1}^\infty (Lu)^{r-1} Q_r.
\]
Here, the combined resolvent $G\indup{mom}(u)$
for momentum-carrying excitations reads
\[
G\indup{mom}(u)=\sum_{k=1}^4\sum_{l=5}^8G_{kl}(u).
\]
Note that we can alternatively write the momentum-carrying resolvent
using the quasi-momenta as
\[\label{eq:Gmom}
G\indup{mom}(u)=\sum_{k=1}^8 \half \sgrad_k \sheetsign_k p_k(u)
=\sum_{k,l=1}^8 \half \sheetsign_k G_{kl}(u).
\]

As usual, the global charges are found in the asymptotics
of the quasi-momenta $\tilde p_k(x)$ at $x=\infty$.
The global charges associated to
$S^5$ at $x=\infty$ are \cite{Beisert:2004ag}
\<\label{eq:SYM.GlobalSO6}
\tilde p_1(u)\eq
\frac{1}{Lu}
\lrbrk{+\sfrac{3}{4}\tilde r_1+\sfrac{1}{2}\tilde r_2+\sfrac{1}{4}\tilde r_3+\sfrac{1}{4}r^\ast}+\cdots\,,
\nln
\tilde p_2(u)\eq
\frac{1}{Lu}
\lrbrk{-\sfrac{1}{4}\tilde r_1+\sfrac{1}{2}\tilde r_2+\sfrac{1}{4}\tilde r_3+\sfrac{1}{4}r^\ast}+\cdots\,,
\nln
\tilde p_3(u)\eq
\frac{1}{Lu} \lrbrk{-\sfrac{1}{4}\tilde r_1-\sfrac{1}{2}\tilde r_2+\sfrac{1}{4}\tilde r_3+\sfrac{1}{4}r^\ast}+\cdots\,,
\nln
\tilde p_4(u)\eq
\frac{1}{Lu}
\lrbrk{-\sfrac{1}{4}\tilde r_1-\sfrac{1}{2}\tilde r_2-\sfrac{3}{4}\tilde r_3+\sfrac{1}{4}r^\ast}+\cdots \,.
\>
Here, $[\tilde r_1,\tilde r_2,\tilde r_3]$ are the the Dynkin labels
of $\alg{su}(4)$ related to orthogonal spins $J_{1,2,3}$ of $\alg{so}(6)$
by $\tilde r_1=J_2-J_3$, $\tilde r_2=J_1-J_2$, $\tilde r_3=J_2+J_3$.
Similarly, in the $\alg{su}(2,2)$ sector \cite{Schafer-Nameki:2004ik}
\<\label{eq:SYM.GlobalSO24}
\hat p_1(u)\eq
-\frac{1}{Lu} \lrbrk{+\sfrac{3}{4}\hat r_1+\sfrac{1}{2}\hat r_2+\sfrac{1}{4}\hat r_3-\sfrac{1}{4}r^\ast}+\cdots\,,
\nln
\hat p_2(u)\eq
-\frac{1}{Lu}
\lrbrk{-\sfrac{1}{4}\hat r_1+\sfrac{1}{2}\hat r_2+\sfrac{1}{4}\hat r_3-\sfrac{1}{4}r^\ast}+\cdots \,,
\nln
\hat p_3(u)\eq
-\frac{1}{Lu} \lrbrk{-\sfrac{1}{4}\hat r_1-\sfrac{1}{2}\hat r_2+\sfrac{1}{4}\hat r_3-\sfrac{1}{4}r^\ast}+\cdots\,,
\nln
\hat p_4(u)\eq
-\frac{1}{Lu}
\lrbrk{-\sfrac{1}{4}\hat r_1-\sfrac{1}{2}\hat r_2-\sfrac{3}{4}\hat r_3-\sfrac{1}{4}r^\ast}+\cdots \,.
\>
The Dynkin labels $[\hat r_1,\hat r_2,\hat r_3]$ of $\alg{su}(2,2)$
are related to orthogonal spins $(S_{1,2},E)$ of $\alg{so}(2,4)$ by $\hat r_1=S_1-S_2$,
$\hat r_2=-E-S_1$, $\hat r_3=S_1+S_2$.
The relation to the fillings reads for $\alg{su}(4)$
\[\label{eq:Moduli.DynkinS5}
\begin{array}{rclcrcl}
\tilde r_1\eq \tilde K_2-2\tilde K_1,&&
\tilde K_1\eq \half L-\sfrac{3}{4}\tilde r_1-\half\tilde r_2-\sfrac{1}{4}\tilde r_3,
\\[0.7ex]
\tilde r_2\eq L-2\tilde K_2+\tilde K_1+\tilde K_3,&&
\tilde K_2\eq \phantom{\half}L-\sfrac{1}{2}\tilde r_1-\phantom{\half}\tilde r_2-\sfrac{1}{2}\tilde r_3,
\\[0.7ex]
\tilde r_3\eq \tilde K_2-2\tilde K_3,&&
\tilde K_3\eq \half L-\sfrac{1}{4}\tilde r_1-\half\tilde r_2-\sfrac{3}{4}\tilde r_3.
\end{array}
\]
and for $\alg{su}(2,2)$ correspondingly
\[\label{eq:Moduli.DynkinAdS5}
\begin{array}{rclcrcl}
\hat r_1\eq \hat K_2-2\hat K_1,&&
\hat K_1\eq -\half L-\half\delta E-\sfrac{3}{4}\hat r_1-\half\hat r_2-\sfrac{1}{4}\hat r_3,
\\[0.7ex]
\hat r_2\eq -L-\delta E-2\hat K_2+\hat K_1+\hat K_3,&&
\hat K_2\eq -\phantom{\half}L-\phantom{\half}\delta E-\sfrac{1}{2}\hat r_1-\phantom{\half}\hat r_2-\sfrac{1}{2}\hat r_3,
\\[0.7ex]
\hat r_3\eq \hat K_2-2\hat K_3,&&
\hat K_3\eq -\half L-\half\delta E-\sfrac{1}{4}\hat r_1-\half\hat r_2-\sfrac{3}{4}\hat r_3.
\end{array}
\]
Note that the Dynkin labels can be obtained as
\<\label{eq:Moduli.DynkinS5Def}
\tilde r_j\eq
\frac{L}{2\pi i}\oint_\infty du\, \bigbrk{\tilde p_j(u)-\tilde p_{j+1}(u)}.
\nln
\hat r_j\eq
\frac{L}{2\pi i}\oint_\infty du\, \bigbrk{\hat p_{j+1}(u)-\hat p_j(u)}.
\>

\subsection{Separation into $\alg{su}(4)$ and $\alg{su}(2,2)$}
\label{sec:Stacks.Separation}

As emphasized above, fermionic stacks are solitary and therefore
irrelevant in the leading approximation of thermodynamic limit. Let
us now write the Bethe equations in a form where the distinction
between fermions and bosons becomes more apparent. We shall
distinguish the quasi-momenta by their grading in the two sets
$\tilde p=\set{p_3,p_4,p_5,p_6}$ corresponding to $\alg{su}(4)$ and
$\hat p=\set{p_1,p_2,p_7,p_8}$ corresponding to $\alg{su}(2,2)$. In
what follows we will denote them as
\[\label{RELABEL}
\begin{array}[b]{rclcrclcrclcrcl}
p_3\earel{\equiv}\tilde p_1,&p_4\earel{\equiv}\tilde p_2,&
p_5\earel{\equiv}\tilde p_3,&p_6\earel{\equiv}\tilde p_4,
\\[4pt]
p_1\earel{\equiv}\hat p_1,&p_2\earel{\equiv}\hat p_2,&
p_7\earel{\equiv}\hat p_3,&p_8\earel{\equiv}\hat p_4.
\end{array}
\]
In the same way we introduce three types of resolvents
$\tilde G_{kl}(u)$ corresponding to bosons of $\alg{su}(4)$,
$\hat G_{kl}(u)$ corresponding to bosons of $\alg{su}(2,2)$
and $G^\ast_{kl}(u)$ corresponding to fermions:
\<
G_{kl}\earel{\equiv} \tilde G_{k'l'},\quad\mbox{for }k,l=3,4,5,6,
\nln
G_{kl}\earel{\equiv} \hat G_{k'l'},\quad\mbox{for }k,l=1,2,7,8,
\nln
G_{kl}\earel{\equiv} G^\ast_{k'l'},\quad\mbox{for }k=3,4,5,6\mbox{ and }l=1,2,7,8,
\>
where the primed indices are related to the unprimed ones
corresponding to \eqref{RELABEL}.
In the leading approximation
of the thermodynamic limit,
the fermionic resolvents are irrelevant, $G^\ast_{kl}=\order{1/L}$, and
we could set them to zero. Let us nevertheless
carry them along to preserve the full supersymmetric structure.

These resolvents obey three sets of independent equations:
\begin{bulletlist}
\item
The Bethe equations for $\alg{su}(4)$ flavor excitations
read
\[
\label{asldkj}
\sheetsl[\tilde]_l(u)-\sheetsl[\tilde]_k(u) =2\pi \tilde n_{kl,a} \qquad \mbox{for }
u\in \tilde{\contour{C}}_{kl,a},\qquad 1\le k<l\le 4
\]
where
\[\label{HATP}
\tilde p_k(u)=
\sum_{l=1}^{4}\tilde G_{kl}(u)
+\sum_{l=1}^{4}G^\ast_{kl}(u)
+\frac{\sheetsign_k}{2u}\,,\qquad k=1,\ldots,4.
\]
Here we have defined the reduced form of $\sheetsign_k$ as
$\sheetsign_k=(+1,+1,-1,-1)$.
Written in terms of resolvents, the Bethe equations read
\[\label{BAEsu4}
\sum_{j=1}^{4}
\bigbrk{-\resolvsl[\tilde]_{kj}(u)+\resolvsl[\tilde]_{lj}(u)-G^\ast_{kj}(u)+G^\ast_{lj}(u)}
-\frac{\sheetsign_k-\sheetsign_l}{2u}=2\pi \tilde n_{kl,a}\quad
\mbox{for }u\in \tilde{\contour{C}}_{kl,a}.
\]

\item
The Bethe equations for $\alg{su}(2,2)$ derivative excitations
read
\[\label{reoityu}
\sheetsl[\hat]_l(u)-\sheetsl[\hat]_k(u) =2\pi \hat n_{kl,a} \qquad
\mbox{for } u\in \hat{\contour{C}}_{kl,a},\qquad 1\le k<l\le 4,
\]
where
\[\label{CHECKP}
\hat p_k(u)=
\sum_{l=1}^{4} \hat G_{lk}(u)
+\sum_{l=1}^{4} G^\ast_{lk}(u)
+\frac{\sheetsign_k}{2u}\,.
\]
Using resolvents, the Bethe equations become
\[\label{BAEsu22}
\sum_{j=1}^{4}
\bigbrk{-\resolvsl[\hat]_{jk}(u)+\resolvsl[\hat]_{jl}(u)-G^\ast_{jk}(u)+G^\ast_{jl}(u)}
-\frac{\sheetsign_k-\sheetsign_l}{2u}=2\pi \hat n_{kl,a}\quad
\mbox{for }u\in \hat{\contour{C}}_{kl,a}.
\]

\item
The Bethe equations for fermionic excitations
read
\[\label{BAEfermi}
\sheetsl[\hat]_l(u)-\sheetsl[\tilde]_k(u) =2\pi n^\ast_{kl,a} \qquad
\mbox{for } u=u^\ast_{kl,a},\qquad 1\le k,l\le 4.
\]
%

\end{bulletlist}
The momentum-carrying resolvent determines the
momentum, energy and local charges of a solution, c.f.~\eqref{ZeroQs},
becomes
\[\label{eq:Stacks.SepMom}
G\indup{mom}
=
\sum_{k,l=1}^4\half\sheetsign_k
\bigbrk{\tilde G_{kl}-\hat G_{lk}+G^\ast_{kl}-G^\ast_{lk}}.
\]

Now let us switch off the fermions by setting $G^\ast_{kl}=0$.
Then we see that the Bethe equations for
the flavor and derivative sectors
\eqref{BAEsu4,BAEsu22} become fully independent.
Moreover these equations are very similar,
they merely differ by the order of indices of resolvents
$\tilde G_{kl}$ and $\hat G_{lk}$. When we order them
in the same way, we see that effectively the
potential term $(\sheetsign_k-\sheetsign_l)/u$ changes sign.
Nevertheless, the two sectors are not completely independent but they are
related through the momentum constraint
$G\indup{mom}(0)=2\pi m$ with $G\indup{mom}(u)$ defined in
\eqref{eq:Stacks.SepMom}. Furthermore, the
energy and local charges are the sums of energy and local
charges in both sectors.

In \secref{sec:scl1} we present an even more particular
case of Bethe equations only for $1+1|4$ modes
in the $\alg{su}(1,1|2)$ sector.

\subsection{Fermions and Solitary Bosons}
\label{sec:Stacks.Solitary}

Let us now discuss the role of fermions in the thermodynamic limit.
As mentioned above, fermionic stacks cannot form strings. The
reason is the exclusion principle for Bethe roots in combination
with the absence of repulsion for fermionic stacks. The absence of
repulsion is the trivial consequence of the absence of
self-scattering terms among fermionic stacks.
In such a fermionic would-be string all the stacks would fall onto the
minimum of the effective potential created by the bare potential
$\pm 1/u+2\pi n$ and the Coulomb interaction with a given
bosonic background. It follows from these arguments that one can
have only solitary fermionic stacks at a given position in the $u$-plane.
Such fermionic excitations have no influence on the
bosonic background in the leading order for $L\to\infty$. The
corresponding resolvents, consisting from a few poles at the
positions of the fermionic stacks, can be dropped from Bethe
equations since the residues at the poles are $\order{1/L}$.
However, in the main order a meaningful problem is to find the
positions of these fermionic poles. In particular, this is useful for
the derivation of the spectrum of excitations of the bosonic
background which is $\order{1/L}$.

The simplest example of such solitary fermionic roots is given by
the purely fermionic $\alg{u}(1|1)$ sector of the $\superN=4$ SYM theory:
The positions $u_n$ of roots are given by the equation
\cite{Callan:2004dt}
\[
\left(\frac{u+\frac{i}{2}}{u-\frac{i}{2}}\right)^L=1,
\]
or in the thermodynamic limit
\[
L/u_n=2\pi n,\qquad n\in \Integers.
\]
Here the effective potential
coincides with the bare one and the fermions sit at its
minima, no more than one in each.

Let us note that we can also consider solitary bosonic
stacks, or very short strings consisting of them. This case
does not differ significantly from the fermionic one: such short
cuts do not contribute in the main order of the thermodynamic
limit and the corresponding resolvents (consisting of a few pole
terms) should be dropped from the Bethe equations.

We can find the positions of such fermionic or solitary bosonic
$(k,l)$-stacks for a bosonic background given by a
solution of \eqref{BAEsu4,BAEsu22,BAEfermi}.
They are the solutions $u$ of the equation
\[
p_l(u)-p_k(u) =2\pi n,
\]
with $p_k(u)$ defined in \eqref{RELABEL,HATP,CHECKP}. Here there is
no principal value prescription since the solitary stack is not
considered part of the quasi-momentum.

For an honest $\order{1/L}$ computation of the energy shift
induced by adding a solitary stack, one must also take into
account the deformation of the bosonic background.
Luckily, the deformation does not influence the leading order position $u$
of the solitary excitation, but it has an effect on the energy
and needs to be investigated.
This is however beyond the scope of the current work.

\subsection{The $\alg{su}(1,1|2)$ Sector}
\label{sec:scl1}

Let us review the above arguments to derive the thermodynamic limit
for the somewhat simpler $\alg{su}(1,1|2)$ Bethe equations
\eqref{b123}. Of course, all the derived equation equally follow by
restriction of the results in the previous subsections. Consider
first $K_v=K_w=K_u-1$, which corresponds to the $\alg{sl}(2)$ sector
after dualization. The middle equation then has a unique solution
which allows one to express $v_p=w_p$'s in terms of $u_q$'s and
reduce the Bethe equations to the $\alg{sl}(2)$ form. How does this
solution look like in the thermodynamic limit? We have to assume
that $v_p\sim \order{L}$, otherwise the energy will be too high.
However, a short inspection of the first equation shows that there
is only finite number of solutions with $v_p\sim \order{L}$
\emph{and} $v_p-u_q=\order{L}$. We have to relax the latter
condition in order to allow for a macroscopically large number of
$v$-roots.  Therefore, the $v$-roots settle close to the $u$-roots
and form stacks in the thermodynamic limit.

\begin{figure}\centering
\includegraphics{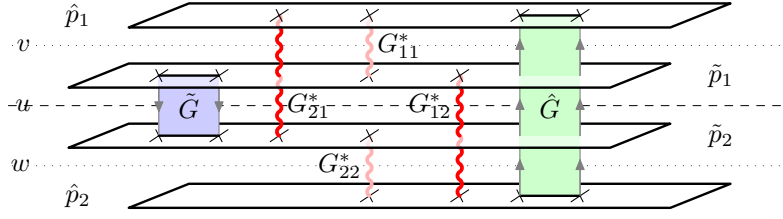}
\caption{
All six types of stacks for the $\alg{su}(1,1|2)$ sector.
The outer four stacks are momentum-carrying,
the inner two are auxiliary.
For each type of stack,
we have indicated the corresponding resolvent.}
\label{fig:dict2}
\end{figure}

In the pure $\alg{sl}(2)$ case all roots (except one) will be
combined into stacks of three: triples $(v_p,u_p,w_p)$,
$p=1,\ldots,K_v=K_w$. Conversely, in the $\alg{su}(2)$ case there is only one
flavor of roots. In a more general situation we have $\hat K$ stacks
of all flavors and $\tilde K$ single roots of flavor $2$,
c.f.~\figref{fig:dict2}.
These two types of stacks are bosonic, are allowed to form strings and we
introduce the corresponding integral resolvents
\[
\hat G(u)=\int_{\tilde{\contour{C}}}
  \frac{dv\,\tilde \rho(v)}{v-u}\,,\qquad
\tilde G(u)=\int_{\hat{\contour{C}}}
  \frac{dv\,\hat \rho(v)}{v-u}\,.
\]
In addition, there can be finitely many fermionic stacks
characterized by the discrete resolvents
with $k,l=1,2$
\[
G^\ast_{kl}(u)=\sum_{p=1}^{K^\ast_{kl}}
  \frac{1/L}{u^\ast_{kl,p}-u}\,.\qquad
\]
We begin with the Bethe equation for roots of flavor $2$.
This is given by the middle equation in \eqref{b123}
and the limit is
\[
2\pi \tilde n_{p}+\frac{L}{u_p}=
\sum_{q=1}^{K_v}\frac{1}{v_q-u_p}
-\sum_{\textstyle\atopfrac{q=1}{q\ne p}}^{K_u}\frac{2}{u_q-u_p}
+\sum_{q=1}^{K_w}\frac{1}{w_q-u_p}.
\]
To turn the sums into the resolvents,
we have to replace a sum over $v_p$ by
$-G^\ast_{21}(u)-G^\ast_{11}(u)+\hat G(u)$,
a sum over $u_p$ becomes
$\tilde G(u)-G^\ast_{21}(u)+G^\ast_{12}(u)+\hat G(u)$
and finally the $w_p$-sums yield
$G^\ast_{22}(u)-G^\ast_{21}(u)+\hat G(u)$,
see~\figref{fig:dict2}.
The resulting Bethe equation reads
\[
2\pi \tilde n_{a}+\frac{1}{u}=
-2\resolvsl[\tilde](u)
-G^\ast_{11}(u)
-G^\ast_{12}(u)
+G^\ast_{21}(u)
+G^\ast_{22}(u)
\quad
\mbox{for }u\in\tilde{\contour{C}}_{a}.
\]
Note that there is no anomaly for a single flavor of roots as
explained in \appref{sec:anom}. Also note that the term $\hat G(u)$
has dropped out from the equations since it has no common label with
$\tilde G(u)$. To obtain the Bethe equation for stacks of three we
first multiply all three Bethe equations and then take the limit
(assuming that a stack is labelled by a common index $p$ for
$u_p\approx v_p\approx w_p$)
\[
2\pi \hat n_{p}+\frac{L}{u_p}=
+\sum_{q=1}^{K_u}\frac{1}{u_q-v_p}
+\sum_{q=1}^{K_v}\frac{1}{v_q-u_p}
-\sum_{\textstyle\atopfrac{q=1}{q\ne p}}^{K_u}\frac{2}{u_q-u_p}
+\sum_{q=1}^{K_w}\frac{1}{w_q-u_p}
+\sum_{q=1}^{K_u}\frac{1}{u_q-w_p}.
\]
In terms of resolvents we obtain
\[
2\pi \hat n_{a}+\frac{1}{u}=
+2\resolvsl[\hat](u)-G^\ast_{11}(u)-G^\ast_{21}(u)+G^\ast_{12}(u)+G^\ast_{22}(u)
\quad
\mbox{for }u\in\hat{\contour{C}}_{a}.
\]
Although the individual Bethe equations are
anomalous, their product is anomaly-free,
c.f.~\appref{sec:anom}.
The equations which determine the
position of the fermionic stacks are determined
in a similar fashion:
\<
2\pi n^\ast_{11,a}\eq
+\tilde G(u)+\hat G(u)-G^\ast_{21}(u)+G^\ast_{12}(u)
\quad
\mbox{for }u=u^\ast_{11,a},
\nln
2\pi n^\ast_{21,a}+\frac{1}{u}\eq
-\tilde G(u)+\hat G(u)-G^\ast_{11}(u)+G^\ast_{22}(u)
\quad
\mbox{for }u=u^\ast_{21,a},
\nln
2\pi n^\ast_{12,a}+\frac{1}{u}\eq
-\tilde G(u)+\hat G(u)-G^\ast_{11}(u)+G^\ast_{22}(u)
\quad
\mbox{for }u=u^\ast_{12,a},
\nln
2\pi n^\ast_{22,a}\eq
+\tilde G(u)+\hat G(u)-G^\ast_{21}(u)+G^\ast_{12}(u)
\quad
\mbox{for }u=u^\ast_{22,a},
\>
In the leading order, we can drop the contribution from the
finitely many fermions. The equations for the remaining bosonic
modes can be written as
\<
-2\pi \tilde n_{a}-\frac{1}{u}\eq
2\resolvsl[\tilde](u)
\quad
\mbox{for }u\in\tilde{\contour{C}}_{a},
\nln
2\pi \hat n_{a}+\frac{1}{u}\eq
2\resolvsl[\hat](u)
\quad
\mbox{for }u\in\hat{\contour{C}}_{a}.
\>
These are two independent equations for $\alg{sl}(2)$ and
$\alg{su}(2)$ densities, which are only coupled through the common
momentum constraint and through the common expression for the
energy
\[
\tilde G(0)+\hat G(0)=2\pi m,
\qquad
\delta D=\frac{\lambda}{8\pi^2L}\,\bigbrk{\tilde G'(0)+\hat G'(0)}.
\]
%

\section{Algebraic Curve}
\label{sec:GaugeCurve}

\begin{figure}\centering
\includegraphics{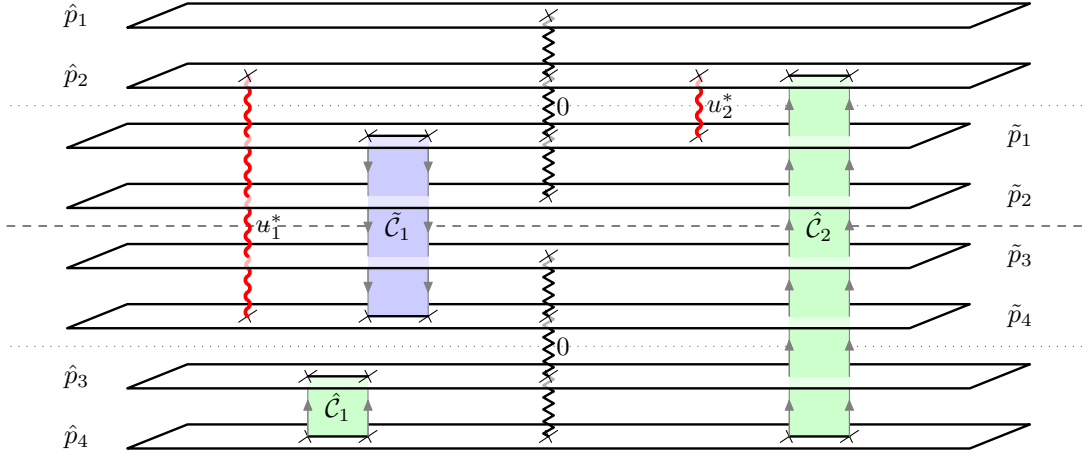}
\caption{Some configuration of cuts and poles for the sigma model.
Cuts $\tilde{\contour{C}}_a$ between the sheets $\tilde p_k$
correspond to $S^5$ excitations and likewise
cuts $\hat{\contour{C}}_a$ between the sheets $\hat p_k$
correspond to $AdS^5$ excitations.
Poles $x^\ast_a$ on sheets $\tilde p_k$ and $\hat p_l$
correspond to fermionic excitations.
The dashed line in the middle is related to physical excitations,
cuts and poles which cross it contribute to the total momentum,
energy shift and local charges.}
\label{fig:sheets}
\end{figure}

We are now in position to construct the complete algebraic curve of
the one-loop $\superN=4$ SYM theory in the thermodynamic limit, in
analogy with the full $AdS_5\times S^5$ curve of the string theory
built in \cite{Beisert:2005bm}. Note that if we drop the fermions,
the problem is not principally new: the bosonic $\alg{su}(4)$ and
$\alg{su}(2,2)$ sectors
are almost completely factorized%
\footnote{They are related only by the global zero momentum
condition \eqref{ZeroQs} enforcing the cyclicity of the trace in the
definition of a SYM operator.}
and each one is described by a separate curve of degree four.

As shown in \cite{Beisert:2004ag},
the $\alg{su}(4)$ spectral curve has the form
\[ \hF(\tilde y,u)=\hF_4(u)\,\tilde y^4 + \hF_2(u)\,\tilde y^2 +
\hF_1(u)\,\tilde y + \hF_0(u)=0
\]
where
\[\label{eq:YDEF}
\tilde y(u)=u^2\,\tilde p'(u).
\]
For a curve with $\tilde A$ branch cuts, the leading coefficient
$\hF_4(u)=\prod_{a=1}^{\tilde A}(u-\tilde u^+_a)(u-\tilde u^-_a)$
encodes the full information about all $2\tilde A$ branch points
$\tilde u^\pm_a$ of the curve. Furthermore, the coefficient of
$y^3$, $\hF_3(u)=0$ due to unimodularity $\tilde p_1(u)+\tilde
p_2(u)+\tilde p_3(u)+\tilde p_4(u)= 0$  of the monodromy matrix in
this sector.
In a $\alg{su}(2,2)$ sector%
\footnote{This sector contains  the scalar field $\mathcal{Z}$ with
arbitrary derivative excitations $\mathcal{D}^k\mathcal{Z}$.} the
situation is similar. The algebraic curve is given by
\cite{Schafer-Nameki:2004ik}
\[
\cF(\hat y,u)=\cF_4(u)\,\hat y^4 + \cF_2(u)\,\hat y^2 +
\cF_1(u)\,\hat y + \cF_0(u)=0,
\]
where again $\cF_4(u)=\prod_{a=1}^{\hat A}(u-\hat u_a^+)(u-\hat u_a^-)$
encodes the full information about the
$2\hat A$ branch points $\hat u^\pm_a$ of the curve.
Note that $\cF_3(u)=0$ since $\hat p_1(u)+\hat p_2(u)+\hat p_3(u)+\hat p_4(u)= 0$.

The supersymmetric curve can be represented, in analogy with the string
algebraic curve \cite{Beisert:2005bm}, in a rational form
reflecting the characteristic \emph{rational function} of a supermatrix
(as opposed to a characteristic \emph{polynomial} of a regular matrix)%
\footnote{The present curve differs from the one
proposed in \cite{Schafer-Nameki:2004ik} by the fact that
it treats the $\alg{so}(4)$- and $\alg{su}(2,2)$-bosons
and the fermions on a different footing (numerator/denominator).
It seems that our degree $4+4$ curve can always be embedded in the
rank $8$ curve of \cite{Schafer-Nameki:2004ik}, but not vice versa.}
\[\label{eq:SPECDET}
F(y,u)=\frac{ \tilde F_4(u)\,y^4 + \tilde F_3(u)\,y^3 + \tilde F_2(u)\,y^2
             +\tilde F_1(u)\,y + \tilde F_0(u)}
            { \hat F_4(u)\,y^4 + \hat F_3(u)\,y^3 +\hat F_2(u)\,y^2
             +\hat F_1(u)\,y +\hat F_0(u)}\,,
\]
with $\tilde F_k(u),\hat F_k(u)$ polynomials in $u$.
The curve $y(u)=\sset{\tilde y(u)}{\hat y(u)}$ obeys the algebraic equations
\[\label{eq:CURVES}
\tilde F(\tilde y(u),u)=0,\qquad \hat F(\hat y(u),u)=0.
\]
Note that now we have
\[\label{FTHREE}
\frac{\hF_3(u)}{\hF_4(u)}=\frac{\cF_3(u)}{\cF_4(u)}
\]
since instead of the two unimodularity conditions we have only one
superunimodularity condition
\[\label{SUNIM}
\tilde p_1(u)+\tilde p_2(u)+\tilde p_3(u)+\tilde p_4(u)
-\hat p_1(u)-\hat p_2(u)-\hat p_3(u)-\hat p_4(u)= 0.
\]
The dynamical singularities are encoded in $F_4(u)$ as
\<\label{CURVEF4} \tilde F_4(u)\eq
\prod_{a=1}^{\tilde A}(u-\tilde u^+_a)
\prod_{a=1}^{\tilde A}(u-\tilde u^-_a)
\prod_{a=1}^{A^\ast}(u-u^\ast_a)^2, \nln
\hat F_4(u)\eq \prod_{a=1}^{\hat A}(u-\hat u^+_a)
\prod_{a=1}^{\hat A}(u-\hat u^-_a)
\prod_{a=1}^{A^\ast}(u-u^\ast_a)^2.
\>
For $\tilde F_4(u),\hat F_4(u)$ there are
in total $2\tilde A+2\hat A+A^\ast$ degrees of freedom.

At $u=\infty$ and $u=0$ the curve $y(u)$ approaches
finite limiting values.
This is achieved by letting all the polynomials
$\tilde F_k(u)$ have the same order $2\tilde A$
as well as all the $\hat F_k(u)$ have the order $2\hat A$.
Note that the behavior of $\hF_{2,1,0},\cF_{2,1,0}$ is generic at
the zeroes of $\hF_4,\cF_4$. The number of free
coefficients in all $\tilde F_k(u),\hat F_k(u)$, $k\le 2$ is
$6\tilde A+6\hat A+12A^\ast+6$.
Adding the parameters of $\tilde F_4(u),\hat F_4(u)$
we get $8\tilde A+8\hat A+13A^\ast+6$ relevant
coefficients.

The unimodularity condition \eqref{FTHREE} imposes
\< \label{CURVEF3}
\tilde F_3(u)\eq F_3^\ast(u)
\prod_{a=1}^{\tilde A}(u-\tilde u^+_a)
\prod_{a=1}^{\tilde A}(u-\tilde u^-_a) , \nln
\hat F_3(u)\eq F_3^\ast(u)
\prod_{a=1}^{\hat A}(u-\hat u^+_a)
\prod_{a=1}^{\hat A}(u-\hat u^-_a) ,
\>
with some polynomial $F_3^\ast(u)$ of order $2A^\ast$,
which contributes $2A^\ast+1$ free coefficients.
In total we now have $8\tilde A+8\hat A+15A^\ast+7$ free parameters.
Let us see how they can be fixed.

\paragraph{Singularities.}

The coefficient of the double pole in
$\tilde p_k',\hat p_k'$
at $u=0$ is read from \eqref{HATP,CHECKP}.
In terms of $y(u)$ it appears as
\[
\tilde y_k(u),\,\hat y_k(u)=-\half \sheetsign_k+\order{u}.
\]
These can be used to fix the relative ratios among
the constant terms $\tilde F_k(0)$
as
$\tilde F_4(0)=-2\tilde F_2(0)=16\tilde F_0(0),\ $
$\tilde F_3(0)=\tilde F_1(0)=0$.
The same constraints hold for $\hat F_k(0)$ except $\hat F_3(0)=0$,
which is automatically satisfied since we have already imposed
\eqref{FTHREE} in the ansatz \eqref{CURVEF3}.
In total we have 7 constraints which reduce the total degrees of freedom
to $8\tilde A+8\hat A+15A^\ast$.

\paragraph{Unphysical Branch Points.}

In addition to the physical branch points at $\tilde u^\pm_a,\hat u^\pm_a$
the algebraic curve might have further ones. Generically,
these singularities are square roots in contrast to the physical ones
which are inverse square roots. We can remove them using conditions
on the discriminants
\<\label{eq:Moduli.Discri}
\tilde R\eq
- 4\tilde F_1^2\tilde F_2^3\tilde F_4
+ 16\tilde F_0\tilde F_2^4\tilde F_4
- 27\tilde F_1^4\tilde F_4^2
+ 144\tilde F_0\tilde F_1^2\tilde F_2\tilde F_4^2
- 128\tilde F_0^2\tilde F_2^2\tilde F_4^2
+ 256\tilde F_0^3\tilde F_4^3
\nl
+ 18\tilde F_1^3\tilde F_2\tilde F_3\tilde F_4
- 80\tilde F_0\tilde F_1\tilde F_2^2\tilde F_3\tilde F_4
- 192\tilde F_0^2\tilde F_1\tilde F_3\tilde F_4^2
- 6\tilde F_0\tilde F_1^2\tilde F_3^2\tilde F_4
+ 144\tilde F_0^2\tilde F_2\tilde F_3^2\tilde F_4
\nl
+ \tilde F_1^2\tilde F_2^2\tilde F_3^2
- 4\tilde F_0\tilde F_2^3\tilde F_3^2
- 4\tilde F_1^3\tilde F_3^3
+ 18\tilde F_0\tilde F_1\tilde F_2\tilde F_3^3
- 27\tilde F_0^2\tilde F_3^4
\>
and similarly for $\hat R$.
The discriminants must have the form
\<\label{eq:SYM.DISCR} \tilde R(u)\eq
\prod_{a=1}^{\tilde A}(u-\tilde u^+_a)
\prod_{a=1}^{\tilde A}(u-\tilde u^-_a) \,\tilde Q(u)^2, \nln
\hat R(u)\eq
\prod_{a=1}^{\hat A}(u-\hat u^+_a)
\prod_{a=1}^{\hat A}(u-\hat u^-_a) \,\hat Q(u)^2.
\>
It is clear that $\tilde u^\pm_a$ and $\hat u^\pm_a$ are roots, because
all terms in \eqref{eq:Moduli.Discri} contain $\tilde F_4$ or
$\tilde F_3$.
The degree of the polynomials $\hat Q,\tilde Q$ is
$5\tilde A+6A^\ast$ and $5\hat A+6A^\ast$, respectively.
This special form of discriminants
removes $5\tilde A+5\hat A+12A^\ast$ parameters
and the remaining number of degrees of freedom is
$3\tilde A+3\hat A+3A^\ast$.

\paragraph{Single Poles and A-Cycles.}

We need to remove all the single poles to avoid undesired
logarithmic behavior in the quasi-momentum.
We can put all the integrals of $dp(u)$ around A-cycles to be zero
\[\label{lkcaxsdl}
\oint_{\contour{A}_a}d\tilde p = 0,
\]
as explained in \cite{Kazakov:2004qf}.
There are $\tilde A+\hat A$ A-cycles around
bosonic cuts.
Fermionic singularities contribute $2A^\ast$
independent single poles: One for $\tilde y$ and one for $\hat y$ at
each $u=u^\ast_a$.
At $u=0$, the curve $y(u)$ has a double pole but no single pole.
This amounts to further $7$ constraints.
Among all these single poles and
A-cycles, there are $7$ relations from the sum over all residues on
all independent sheets.
In total the single-valuedness of $p(u)$ yields
$\tilde A+\hat A+2A^\ast$ constraints and leaves
$2\tilde A+2\hat A+A^\ast$ coefficients.

\paragraph{B-Periods.}

For each bosonic cut and for each fermionic singularity there is a
B-period.
Due to the Bethe equations \eqref{asldkj,reoityu,BAEfermi}
it must be an integral multiple of $2\pi$
\[\label{amcsdf}
\int_{\contour{B}_a}dp=2\pi n_a.
\]
In total we obtain $\tilde A+\hat A+A^\ast$
constraints and are left with $\tilde A+\hat A$ degrees of
freedom.

\paragraph{Cyclicity.}

Finally we should not forget the momentum constraint
in \eqref{ZeroQs}
\[\label{sdacwlqek}
q(0)=2\pi m,
\qquad q(u)=
\sum_{k=1}^4\sheetsign_k
\bigbrk{\tilde p_k(u)-\hat p_k(u)}
=G\indup{mom}(u)
\]
selecting curves with an integer winding number $m$.
In total the admissible algebraic curves
for one-loop gauge theory have $\tilde A+\hat A-1$
continuous moduli.

Let us just repeat \eqref{ZeroQs}, which defines
the energy shift $\delta E$ and local charges $Q_r$,
in terms of the generating function $q(u)$
\[\label{asdlfkajsd}
\delta E=\frac{\lambda}{8\pi^2 L}\, q'(0).
\qquad
q(u)=\sum_{r=1}^\infty (Lu)^{r-1} Q_r.
\]

\paragraph{Fillings.}

We can now associate a particular integral around a cut,
the \emph{filling}, to each of the moduli
\<
\tilde K_{kl,a}\eq-\frac{L}{2\pi i}\oint_{\tilde{\contour{A}}_{kl,a}}
\tilde p_k(u)\,du,
\nln
\hat K_{kl,a}\eq-\frac{L}{2\pi i}\oint_{\hat{\contour{A}}_{kl,a}}
\hat p_l(u)\,du,
\nln
K^\ast_{kl,a}\eq-\frac{L}{2\pi i}\oint_{{\contour{A}}^\ast_{kl,a}}
\tilde p_k(u)\,du.
\>
For completeness, we have defined fillings for fermionic poles as well,
they must equal $1$.
The fillings obey the constraint
\[\label{LENGTH}
\sum_{a=1}^A n_a K_a=m L
\]
which can be derived from the Bethe equations
\eqref{BAEfull}
by integrating/summing over all cuts/poles.
This reduces the number of independent
fillings by one, in compliance with
the total number of independent moduli.

\paragraph{Supersymmetric Landau-Lifshitz Model.}

We would like to note
that the curve discussed in this section is
most likely the spectral curve of the
generalized Landau-Lifshitz model
proposed in \cite{Mikhailov:2004xw}.%
\footnote{We thank A.~Mikhailov and R.~Roiban for discussions of
this issue. Supersymmetric Landau-Lifshitz models are also discussed
in \cite{Hernandez:2004kr,Bellucci:2005vq,Stefanski:2005tr}.} This
model appears to arise as the coherent-state picture in the
thermodynamic limit of one-loop gauge theory. It seems to be a
rather straight-forward generalization of the
$\grp{SU}(2)/\grp{U}(1)$ coherent-state model in
\cite{Kruczenski:2003gt} to the larger and supersymmetric coset
$\grp{PSU}(2,2|4)/\grp{PSU}(2|2)^2\times\grp{U}(1)^2$. For the
smaller model, the Heisenberg magnet, the finite-gap solution was
constructed in \cite{Kazakov:2004qf}. Like the above curve, this one
has only one singularity at $u=0$ and $\alg{psu}(2,2|4)$ symmetry
generators at $u=\infty$. Furthermore, the highest weight state
$\fldZ$ is invariant under $\grp{PSU}(2|2)^2$ and thus the residues
$\sheetsign_k/2$ at $u=0$ should match. It would be interesting to
construct some local charges of the supersymmetric Landau-Lifshitz
model which arise in the expansion around $u=0$.

\section{Comparison to String Theory}
\label{sec:Compare}

\subsection{Spectral Curves}
\label{sec:Compare.Curve}

Now we will show that the most general algebraic curve of the SYM
theory at one loop, constructed in the last section,
coincides with that of the string sigma model
\cite{Beisert:2005bm} in the Frolov--Tseytlin
limit $\lambda/L^2\to 0$ (see Sec.~3.7 of \cite{Beisert:2005bm}).
This coincidence was already observed for particular sectors in
\cite{Kazakov:2004qf,Kazakov:2004nh,Beisert:2004ag}.

As in \cite{Beisert:2004ag} we compare the analytical data defining
the structure and the moduli of both curves in terms of a rescaled
variable $u$ which is defined as $u=(\sqrt{\lambda}/4\pi L)\,x$ for
the strings and is  similar to the spectral parameter $u$ used
through this paper.
The parameter of ``length{}'' $L$ for strings was introduced
in Sec.~3.3 of \cite{Beisert:2005bm}
as a modulus satisfying the equivalent of the constraint \eqref{LENGTH}.

The algebraic curve for the string sigma model exhibits
an inversion symmetry $x\mapsto 1/x$ with respect to the coordinate $x$
and all the branch points and fermionic
poles form mutually symmetric pairs.%
\footnote{The case of self-symmetric cuts apparently does not
lead to a proper integer-power expansion in the Frolov--Tseytlin limit.}
In terms of the coordinate $u$ they can be expressed as
\[
(u_a,u_{A+a})=\left(u_a,\frac{\lambda}{16\pi^2 L^2}\,\frac{1}{u_a}\right),
\quad a=1,\ldots,A
\]
where we abbreviate $\tilde u_a^\pm,\tilde A$,
$\hat u_a^\pm,\hat A$ or $u_a^*,A^*$ by $u_a,A$.
In the Frolov--Tseytlin limit, the $u_a$'s remain finite while
the $u_{A+a}$'s go to zero.
This means that in the $x$-plane half of
the cuts/poles approach $x=\infty$
and half of them approach $x=0$.
We will concentrate our attention on that half of the cuts/poles which remain
finite in the $u$-plane: $\tilde{\contour{C}}_a,\ a=1,\ldots,\tilde A$,
$\hat{\contour{C}}_a,\ a=1,\ldots,\hat A$ and $u_a^*,\ a=1,\ldots,A^*$.
The other half of the cuts become infinitely short in the limit and the
corresponding behavior at $u=0$ needs to be treated separately
by completing the analytic structure of the curve at this point.
We are thus left with half of the cuts/poles having no symmetry with
respect to inversion $x\to 1/x$, as in the case of the SYM curve.

Both curves have the following common properties:
\begin{bulletlist}
\item
In the limit $L\to\infty$
the overall structure of the spectral curve
of the sigma model in Sec.~3.2 of \cite{Beisert:2005bm}
becomes the same as in \secref{sec:GaugeCurve}
of the present paper for similar definitions of
$\tilde y(u),\hat y(u)$.

\item
Any two of the eight sheets $y(u)$ of the curve can be connected by
cuts/poles in both sigma model and gauge theory.
Note however that in gauge theory the cuts connecting the non-neighboring
sheets, say sheet 2 and sheet 8, are possible only due to the
existence of stacks introduced in \secref{sec:Stacks}.

\item
The same conditions \eqref{lkcaxsdl,amcsdf}
of zero A-cycles and integrality of B-cycles
for cuts and poles as in Sec.~3.1 of \cite{Beisert:2005bm}
for the sigma model.

\item
The same asymptotics for $p(u)$ at $u\to\infty$ for all the sheets,
given through the $\grp{SO}(6)$ and $\grp{SU}(2,4)$ charges
in Sec.~2.8 of \cite{Beisert:2005bm}
for the sigma model and by the equations
\eqref{eq:SYM.GlobalSO6,eq:SYM.GlobalSO24} of the present
paper for the SYM theory.

\item
The singular behavior of $p(u)$ at $u=0$
for the sigma model in the Frolov--Tseytlin limit
is studied in Sec.~3.7 of \cite{Beisert:2005bm}.
After fixing a gauge for the $\grp{U}(1)$ hypercharge $B=0$ of the sigma model,
$p(u)$ exhibits the same pole structure
$\tilde p_k(u)\sim \hat p_k(u)\sim \sheetsign_k/2u$
as that of gauge theory following from the definitions \eqref{HATP,CHECKP}.
The extra poles at zero for the sigma model come from the poles
at $u=\pm \sqrt{\lambda}/(4\pi L)$ when $\lambda/L^2\to 0$.

\item
The generating function $q$ of local charges of the string
from Sec.~2.6,~3.7 in \cite{Beisert:2005bm}
coincides in the limit with $q$ of gauge theory defined in \eqref{sdacwlqek}
and hence generates the same local charges \eqref{asdlfkajsd}.
In particular, the momentum constraint
\eqref{sdacwlqek} and definition of energy shift \eqref{asdlfkajsd} are the same.

\end{bulletlist}

These properties define the one-loop algebraic curves and their
relation to the physical data unambiguously and consequently they
coincide.

\subsection{Bethe Equations}
\label{sec:Compare.Bethe}

Here we shall compare the thermodynamic limit of the Bethe equations
\eqref{asldkj,reoityu} to the integral equations that describe
classical superstrings in $AdS_5\times S^5$. The integral equations
for the spectral data of the string have the same form as
\eqref{asldkj,reoityu}, but the quasi-momenta are defined a little
differently \cite{Beisert:2005bm}:
\<\label{bestring}
 \tilde{p}_k(u)\eq\sum_{l=1}^{4}
\bigbrk{\tilde{G}_{kl}(u)+\tilde{G}^\ast_{kl}(u)}
+\sheetsign_k\tilde{F}(u)+F^\ast(u),
\nln
 \hat{p}_k(u)\eq\sum_{l=1}^{4}
\bigbrk{\hat{G}_{lk}(u)+\tilde{G}^\ast_{lk}(u)}
+\sheetsign_k\hat{F}(u)+F^\ast(u).
\>
The spectral parameters in the gauge and string theory are
non-trivial functions of one another: $x+1/x=4\pi
Lu/\sqrt{\lambda}$, but at one loop only the linear part of the
transformation matters: $x=4\pi Lu/\sqrt{\lambda}+\cdots$. The
forces (as functions of the string spectral parameter) are given by
\<\label{eq:Integral.Potentials} \tilde F(x)\eq \lrbrk{\frac{2\pi\,
L}{\sqrt{\lambda}}+\tilde G'\indup{sum}(0)} \frac{1/x}{1-1/x^2}
+\frac{\hat G\indup{sum}(0)}{1-1/x^2} -\tilde G\indup{sum}(1/x)
+\tilde{G}\indup{sum}(0)-\hat{G}\indup{sum}(0) , \nln \hat F(x)\eq
\lrbrk{\frac{2\pi\, L}{\sqrt{\lambda}}+\tilde G'\indup{sum}(0)}
\frac{1/x}{1-1/x^2} +\frac{\hat G\indup{sum}(0)}{1-1/x^2} -\hat
G\indup{sum}(1/x), \nln F^\ast(x)\eq
\lrbrk{\frac{2\pi\,B}{\sqrt{\lambda}}+G^{\ast\prime}\indup{sum}(0)}\frac{1/x}{1-1/x^2}
+\frac{G^\ast\indup{sum}(0)}{1-1/x^2} -G^\ast\indup{sum}(1/x) \,. \>
with some combinations $G\indup{sum}$ of the resolvents which we do
not have to specify any closer. To recover \eqref{BAEsu4,BAEsu22},
we expand in $\lambda$ to the leading order. The external force
simplifies in that limit to $\tilde{F}(u)=1/2u=\hat{F}(u)$ and
$F^\ast(u)=0$ when we fix the hypercharge of the vacuum $B=0$.

When, at leading order, we consistently drop the fermions in $G^\ast$,
the crucial point is the separation of
the $\alg{su}(4)$ and $\alg{su}(2,2)$ excitations via stacks.
Equations \eqref{asldkj} and \eqref{reoityu} become independent.
Their solutions are only connected through the common
expressions for the momentum and energy.
This is precisely what we expect from the bosonic string where
the motion of the string in $S^5$ and in $AdS_5$ is independent
and only related by the Virasoro constraint.

\section{Conclusions}
\label{sec:Conclusions}

We have studied the spectrum of gauge-invariant local single-trace
operators in the ${\cal N}=4$ super Yang-Mills theory at one loop.
The emphasis is on the so-called thermodynamic limit of long
operators, consisting of many elementary fields. This limit can be
compared directly with a certain classical limit for the dual
string theory on $AdS_5\times S^5$ background. The full classical
finite-gap solution of this integrable string theory was presented in
our previous paper \cite{Beisert:2005bm}.

As is now well established (see \cite{Beisert:2004ry} and references therein),
the analysis of one-loop planar $\superN=4$ SYM theory can be greatly
simplified due to integrability and the spectral problem can be
viewed as a set of algebraic equations on a set of Bethe roots arising from the
Bethe ansatz description of the $\alg{su}(2,2|4)$ quantum spin chain.
These equations can be viewed as an electrostatic
equilibrium problem for 2D Coulomb charges with coordinates given
by the roots.

After presenting the details of the Bethe ansatz description, we
have discussed a duality transformation, which appears to be a crucial
tool for the unified understanding of Bethe equations for super Lie algebras:
A super spin chain admits various different sets
of Bethe equations as fundamental equations due to the
non-uniqueness of the assignment of the gradings in the super Lie
algebra. The duality transformation connects all such descriptions
and is accompanied by the introduction of dual Bethe roots. It
serves as a transformation not only for the Bethe equations but
also for local charges such as the momentum and the energy. We have
presented the most general transformation for the $\alg{sl}(m|n)$
algebras and then illustrated it for the $\alg{su}(2,2|4)$ algebra
as well as for the $\alg{su}(1,1|2)$ subsector.

Another key observation of our analysis was the formation of stacks
of Bethe roots in the thermodynamic limit. The existence of stacks
is crucial to complete the precise gauge/string dictionary.
A stack is a bound state of Bethe roots of different
flavors (nodes of the Dynkin diagram) and is formed due to the
attractive force among them. Algebraically, the stacks represent
positive roots in the same way as single Bethe roots represent
simple roots of the $\alg{su}(2,2|4)$ algebra. On the string
side, stacks correspond to elementary excitations of the world sheet.
In this respect it would be interesting to understand what are the
collective ``coordinates{}'' of stacks on the spectral plane before the
thermodynamic limit. A hint may come from the duality
transformations of the Bethe equations between different systems of roots of the
superalgebra, where apparently the dual roots play the role of such
``coordinates{}'', at least for some types of stacks. Another question,
interesting for the theory of the Bethe ansatz: Do the stacks
survive in the more traditional thermodynamic limit when the length
of the spin chain is large but the Bethe roots
remain finite instead of scaling out to infinity?

Finally, we solve the Bethe equations in the thermodynamic limit
by constructing an algebraic curve which contains all
the relevant information on the anomalous dimension of the
corresponding long SYM operator.
The curve perfectly agrees with the one of the
classical sigma model on $AdS_5\times S^5$ in the Frolov--Tseytlin limit
which was constructed in \cite{Beisert:2005bm}.
Various strings of stacks form cuts which may connect any two remote
Riemann sheets, in the same way as strings of single roots form cuts
between two adjacent Riemann sheets. Stacks corresponding to
fermionic roots of Lie superalgebras stay solitarily and do not form
cuts, in similarity with the fermionic poles found on the sigma model
side \cite{Beisert:2005bm}.
Note that in the thermodynamic limit,
the duality transformation reduces to nothing
but a relabelling of sheets of the Riemann surface of the curve.

We hope that the understanding of basic properties of the Bethe
equations diagonalizing the one-loop dilatation operator of SYM
theory can play a significant role in advancing us to the main
purpose: discovery of the complete integrable structure of the
theory for arbitrary strength of the coupling and extracting some
non-trivial physical quantities out of it. One problem which seems
to be at the reach of our present possibilities is the
generalization of the one-loop Bethe equations to two and three
loops, as it was already done in a the smallest closed sectors
$\alg{su}(2)$ \cite{Serban:2004jf} as well as $\alg{u}(1|1)$ and
$\alg{sl}(2)$ \cite{Staudacher:2004tk}, or even asymptotically in
$\alg{su}(2)$ \cite{Beisert:2004hm}. Our previous paper
\cite{Beisert:2005bm} provides valuable information for this purpose
since it allows to compare the classical solution of the full dual
string theory with the thermodynamic limit of the SYM spin chain.
The higher order corrections on both sides can be also compared,
though we know that some discrepancies, attributed to differences of
weak and strong coupling regime, start to appear already from the
third order \cite{Callan:2003xr,Serban:2004jf}.

Maybe the most promising prospect stemming  from our  results of
this and  the previous paper \cite{Beisert:2005bm} is to attempt
quantizing the full string theory directly and thus to construct the
solution of both sides of the AdS/CFT correspondence. In contrast to
the $\Real_t\times S^3$ sigma model, corresponding to the
$\alg{su}(2)$ sector of SYM theory,%
\footnote{It was argued in \cite{Minahan:2005jq} that a rigorous
restriction to the $\alg{su}(2)$ subsector is
impossible on the string side of AdS/CFT.
The $\Real_t\times S^3$ sigma model
corresponds to a restriction of gauge theory to
$\alg{so}(4)=\alg{su}(2)+\alg{su}(2)$ states.}
we know that the full
superstring sigma model on $AdS_5\times S^5$ is conformally
invariant and can possibly be quantized by transforming to
action/angle variables. As we noticed in \cite{Beisert:2005bm}, the
algebraic curve established there provides the action variables and
identifies in principle the angle variables to be quantized.

In spite of spectacular progress of the last years, the subject of
the AdS/CFT correspondence and the attempts to find its full
integrable structure are still at the ``experimental{}'' stage
 \cite{Polyakov:2004qq}.
We are collecting and comparing various facts on both sides of
AdS/CFT integrability. We do not know for sure which of them will
lead  to the  solution of this important problem,  but we hope that
the structure of the complete theory (and not only of its particular
sectors) unveiled here and in \cite{Beisert:2005bm}  will
substantially improve our ``experimental{}'' tools.

\subsection*{Acknowledgements}

We thank V.~Bazhanov, A.~Gorsky, L.~Lipatov, A.~Mikhailov,
J.~Minahan, R.~Roiban, D.~Serban, F.~Smirnov, A.~Sorin,
M.~Staudacher, A.~Tseytlin, P.~Wiegmann, A.~Zamolodchikov for
discussions and useful comments on the manuscript. N.~B.~would like
to thank the Ecole Normale Sup\'erieure for the kind hospitality
during parts of this work. The work of N.~B.~is supported in part by
the U.S.~National Science Foundation Grant No.~PHY02-43680. Any
opinions, findings and conclusions or recommendations expressed in
this material are those of the authors and do not necessarily
reflect the views of the National Science Foundation.  The work of
V.~K.~was partially supported by the European Union under the RTN
contracts HPRN-CT-2000-00122 and 00131 and by NATO grant
PST.CLG.978817. The work of K.~S.~is supported by the Nishina
Memorial Foundation. The work of K.~Z. was supported in part by the
Swedish Research Council under contracts 621-2002-3920 and
621-2004-3178, by G\"oran Gustafsson Foundation and by RFBR grants
02-02-17260 and NSh-1999.2003.2.

\appendix



\section{Sleeping Beauty}
\label{sec:Beauty}

This appendix contains lengthy expressions related
to the complete superalgebra
using the `Beauty' form of $\alg{su}(2,2|4)$ \cite{Beisert:2003yb},
c.f.~\figref{fig:Beauty}.
In this form, the grading of the sheets corresponding
to the fundamental representation reads
\[
\sgrad_k=(-1,-1,+1,+1,+1,+1,-1,-1).
\]
The sheets of the quasi-momentum are arranged
according to \eqref{RELABEL}
\[\label{eq:Moduli.BeautySheets}
p_{1,2,7,8}=\hat p_{1,2,3,4},\qquad
p_{3,4,5,6}=\tilde p_{1,2,3,4}.
\]
%

\subsection{Bethe Equations}
\label{sec:Beauty.Gauge}

Written out explicitly, the one-loop Bethe equations
\eqref{eq:Gauge.Bethe} for gauge theory read
\<\label{eq:Beauty.GaugeBethe}
1\eq
\phantom{\prod\nolimits_q
 s_{+1}\bigbrk{u^{(1)}_p-u^{(2)}_q}}
\prod\nolimits_q^{\prime}
 s_{-2}\bigbrk{u^{(1)}_p-u^{(1)}_q}
\prod\nolimits_q
 s_{+1}\bigbrk{u^{(1)}_p-u^{(2)}_q},
\nln
1\eq
\prod\nolimits_q
 s_{+1}\bigbrk{u^{(2)}_p-u^{(1)}_q}
\,\phantom{\prod\nolimits_q^{\prime}
 s_{-2}\bigbrk{u^{(1)}_p-u^{(1)}_q}}
\prod\nolimits_q
 s_{-1}\bigbrk{u^{(2)}_p-u^{(3)}_q},
\nln
1\eq
\prod\nolimits_q
 s_{-1}\bigbrk{u^{(3)}_p-u^{(2)}_q}
\prod\nolimits_q^{\prime}
 s_{+2}\bigbrk{u^{(3)}_p-u^{(3)}_q}
\prod\nolimits_q
 s_{-1}\bigbrk{u^{(3)}_p-u^{(4)}_q},
\nln
\Bigbrk{s_{+1}\bigbrk{u_p^{(4)}}}^L \eq
\prod\nolimits_q
 s_{-1}\bigbrk{u^{(4)}_p-u^{(3)}_q}
\prod\nolimits_q^{\prime}
 s_{+2}\bigbrk{u^{(4)}_p-u^{(4)}_q}
\prod\nolimits_q
 s_{-1}\bigbrk{u^{(4)}_p-u^{(5)}_q},
\nln
1\eq
\prod\nolimits_q
 s_{-1}\bigbrk{u^{(5)}_p-u^{(4)}_q}
\prod\nolimits_q^{\prime}
 s_{+2}\bigbrk{u^{(5)}_p-u^{(5)}_q}
\prod\nolimits_q
 s_{-1}\bigbrk{u^{(5)}_p-u^{(6)}_q},
\nln
1\eq
\prod\nolimits_q
 s_{-1}\bigbrk{u^{(6)}_p-u^{(5)}_q}
\,\phantom{\prod\nolimits_q^{\prime}
 s_{-2}\bigbrk{u^{(7)}_p-u^{(7)}_q}}
\prod\nolimits_q
 s_{+1}\bigbrk{u^{(6)}_p-u^{(7)}_q},
\nln
1\eq
\prod\nolimits_q
 s_{+1}\bigbrk{u^{(7)}_p-u^{(6)}_q}
\prod\nolimits_q^{\prime}
 s_{-2}\bigbrk{u^{(7)}_p-u^{(7)}_q}
\>
with the scattering term
\[\label{eq:Beauty.GaugeScatter}
s_t(u)=\frac{u+\sfrac{i}{2}t}{u-\sfrac{i}{2}t}\,.
\]
The products go over all allowed values $q=1,\ldots,K_j$ for the
particular flavor $j$ of Bethe roots $u^{(j)}_q$.
For the primed product means that
that the term with coinciding roots, $q=p$, is omitted.

\subsection{Global Charges}
\label{sec:Beauty.Charges}

The excitation numbers $K_j$ are related to the Dynkin labels
$[r_1;r_2;r_3,r_4,r_5;r_6;r_7]$ of the state through
\<\label{eq:Beauty.DynkinFilling}
r_1\eq K_2-2K_1,
\nln
r_2\eq K_3-K_1+\half \delta E,
\nln
r_3\eq K_2+K_4-2K_3,
\nln
r_4\eq L-2K_4+K_3+K_5,
\nln
r_5\eq K_4+K_6-2K_5,
\nln
r_6\eq K_5-K_7+\half \delta E,
\nln
r_7\eq K_6-2K_7.
\>
or for short
\[\label{eq:Beauty.DynkinFillingShort}
\bar\sgrad_j r_j=V_j\, L+\bar V_j\,\delta E-M_{jj'}K_{j'}.
\]
Here $\bar\sgrad_j=[-1;+1;+1,+1,+1;+1;-1]$ are
conventional factors for the definition of the Dynkin labels.
The Dynkin labels $V_j=[0;0;0,1,0;0;0]$ characterize the spin representation
to which each field of $\superN=4$ SYM belongs.
The labels $\bar V_j=[0;\half;0,0,0;\half;0]$ indicate the
change of (fermionic) Dynkin labels induced by the energy shift.
Note that the Dynkin labels obey the central charge constraint
\[\label{eq:Beauty.DynkinCentral}
-r_1+2r_2+r_3=r_5+2r_6-r_7.
\]
The inverse relation between Dynkin labels and excitation numbers is given by
\[\label{eq:Beauty.FillingDynkin}
\arraycolsep1.2pt
\begin{array}{rccrcrcrcrcrcrcrcrcrcrcr}
K_1&=&-&\half L&+&\half B &-&\sfrac{1}{4}r^\ast&-&\sfrac{3}{4} r_1&+&\half r_2&+&\half r_3&+&\half r_4&+&\half r_5&+&\half r_6&-&\sfrac{1}{4} r_7&-&\half \delta E,
\\[0.7ex]
K_2&=&-&L&+&B&-&\half r^\ast& -&\half r_1&+&r_2&+&r_3&+&r_4&+&r_5&+&r_6&-&\sfrac{1}{2} r_7&-&\delta E,
\\[0.7ex]
K_3&=&-&\half L&+&\half B&-&\sfrac{1}{4}r^\ast& -& \half r_1 &+& r_2& +&\sfrac{1}{4}r_3 &+&\half r_4&+&\sfrac{3}{4}r_5&+&r_6&-& \half r_7&-&\delta E,
\\[0.7ex]
K_4&=&&&&&&& -& \half r_1 &+& r_2 &+& \half r_3&&& +& \half r_5&+&r_6&-&\half r_7 &-&\delta E,
\\[0.7ex]
K_5&=&-&\half L&-&\half B&+&\sfrac{1}{4}r^\ast&-&\half r_1&+&r_2&+&\sfrac{3}{4}r_3&+&\half r_4&+&\sfrac{1}{4} r_5&+&r_6&-&\sfrac{1}{2} r_7& -& \delta E,
\\[0.7ex]
K_6&=&-&L&-&B &+&\half r^\ast& -&\half r_1&+&r_2&+&r_3&+&r_4&+&r_5&+&r_6&-&\half r_7&-&\delta E ,
\\[0.7ex]
K_7&=&-&\half L&-&\half B &+&\sfrac{1}{4}r^\ast&-&\sfrac{1}{4} r_1&+&\half r_2&+&\half r_3&+&\half r_4&+&\half r_5&+&\half r_6&-&\sfrac{3}{4} r_7&-&\half\delta E.
\\[0.7ex]
\end{array}
\]
The constant $B$ represents the hypercharge of the vacuum.
In gauge theory we conventionally set it to zero.
The label $r^\ast$ describes the hypercharge of the state.
It is a conserved quantity at the one-loop level, but
violated by higher-loop effects.

In the thermodynamic limit, the excitation numbers
are replaced by the global fillings defined as
\[\label{eq:Beauty.FillingsGauge}
K_j=
\sum_{a=1}^{A}
\frac{L}{2\pi i}\oint_{\contour{C}_a} du
\sum_{k=1}^j \sgrad_kp_k(u).
\]
The global filling $K_j$ essentially measures the total filling of all
$(k,l)$-cuts with $k\leq j< l$.
The Dynkin labels are directly obtained through the residues at infinity
\[\label{eq:Beauty.DynkinGauge}
\bar\sgrad_j r_j=
\frac{L}{2\pi i}\oint_\infty du\, \bigbrk{p_{j}(u)-p_{j+1}(u)}.
\]
%

\subsection{Dualization of Excitation Numbers}

Let $\{K_j\}_{j=1}^7$, $\{\tK_j\}_{j=1}^7$ respectively
denote the excitation numbers in the Beauty and Beast basis.
According to \eqref{eq:BNTrf} and \eqref{eq:BNTrfs0} one obtains
\[\label{ENumBy2Bt}
\begin{array}[b]{ccr@{\,}r@{\,}r@{\,}r@{\,}r@{\,}r@{\,}r@{\,}r@{\,}r}
\tK_1\Eqn{=}  &K_1,&     &    &   &   &    &    &   \\[0.3ex]
\tK_2\Eqn{=}  &    & K_2,\!\!&&   &   &    &    &   \\[0.3ex]
\tK_3\Eqn{=}2L&    &+K_2 &    &   &   &-K_6&+K_7&-4,\\[0.3ex]
\tK_4\Eqn{=}2L&    &+K_2 &    &   &   &-K_6&    &-8,\\[0.3ex]
\tK_5\Eqn{=} L&    &     &+K_3&   &   &-K_6&    &-6,\\[0.3ex]
\tK_6\Eqn{=}  &    &     &    &K_4&   &-K_6&    &-4,\\[0.3ex]
\tK_7\Eqn{=}  &    &     &    &   &K_5&-K_6&    &-2.
\end{array}
\]
The representation of the state
does not change under dualization.
However, the interpretation of highest weight
state changes by
$\tilde E=E+4$, $\tilde r^\ast=r^\ast+4$,
c.f.~\appref{sec:Beauty.Charges},
and leads to the constant finite shifts
in \eqref{ENumBy2Bt}.%
\footnote{This is for generic long multiplets with
$0\le K_1\le K_2\le K_3-2\le K_4-4\ge K_5-2\ge K_6\ge K_7\ge 0$
and $K_2+2L\ge K_3+L+2\ge K_4+4\le K_5+L+2\le K_6+2L$.
Short multiplets lead to special cases in the
duality transformation and the excitation numbers are slightly modified.}
The remaining Beauty labels $r_1,r_3,r_4,r_5,r_7$ do not change,
c.f.~\cite{Beisert:2003yb} for the translation of $\tilde K_j$ into
Dynkin labels.
The inverse relation reads
\[\label{ENumBt2By}
\begin{array}[b]{ccr@{\,}c@{\,}c@{\,}c@{\,}c@{\,}c@{\,}c@{\,}c@{\,}c}
K_1\Eqn{=}  & \tK_1,&     &   &    &    &    &    &   \\[0.3ex]
K_2\Eqn{=}  &     & \tK_2,\!\!\!\!\!\!&&&    &    &   \\[0.3ex]
K_3\Eqn{=} L&     &+\tK_2 &   &-\tK_4&+\tK_5&    &    &-2,\\[0.3ex]
K_4\Eqn{=}2L&     &+\tK_2 &   &-\tK_4&    &+\tK_6&    &-4,\\[0.3ex]
K_5\Eqn{=}2L&     &+\tK_2 &   &-\tK_4&    &    &+\tK_7&-6,\\[0.3ex]
K_6\Eqn{=}2L&     &+\tK_2 &   &-\tK_4&    &    &    &-8,\\[0.3ex]
K_7\Eqn{=}  &     &     &\tK_3&-\tK_4&    &    &    &-4.
\end{array}
\]

For the interchange of
$\alg{su}(4)$ and $\alg{su}(2,2)$, c.f.~\figref{fig:su2sl},
the excitation numbers are related by
\[
\begin{array}[b]{ccr@{\,}c@{\,}c@{\,}c@{\,}c@{\,}c@{\,}c@{\,}c@{\,}c}
\tK_1\Eqn{=}   &-K_2&+K_3&     &    &    &    &-2,\\[0.3ex]
\tK_2\Eqn{=}   &-K_2&    &+K_4 &    &    &    &-4,\\[0.3ex]
\tK_3\Eqn{=}K_1&-K_2&    &+K_4 &    &    &    &-2,\\[0.3ex]
\tK_4\Eqn{=}   &    &    & K_4,\!\!\!\!\!\!&    &    &    &   \\[0.3ex]
\tK_5\Eqn{=}   &    &    & K_4\!\!\!\!&    &-K_6&+K_7&-2,\\[0.3ex]
\tK_6\Eqn{=}   &    &    & K_4\!\!\!\!&    &-K_6&    &-4,\\[0.3ex]
\tK_7\Eqn{=}   &    &    &     & K_5&-K_6&    &-2.
\end{array}
\]
Here the highest weight state is shifted by
$\tilde E=E+4$, $\tilde J=J+4$,
with $J=\half r_3+r_4+\half r_5$ the main spin of $\alg{su}(4)$
in the Beauty basis.
All the other Beauty labels $r_1,r_3,r_5,r_7$ remain unchanged.

\subsection{Fundamental Transfer Matrix}
\label{sec:Beauty.Transfer}

A central object of the Bethe ansatz is a formula which determines
the eigenvalue of a transfer matrix%
\footnote{The transfer matrices
for the $\alg{sl}(n|m)$ Bethe ansatz are discussed in
\cite{Tsuboi:1998ne,Tsuboi:1997iq,Tsuboi:1998sc}.} 
for a given set of Bethe roots $\set{u_p^{(j)}}$. 
The transfer matrix in the
fundamental representation is \cite{Schafer-Nameki:2004ik}
\[\label{eq:TMBeauty}
\begin{array}[b]{r@{\,}l@{\;}l@{\;}l}
T_{\rep{4|4}}(u)
=-&S_1^+(u)  &                     &U_{+1}(u)\\[0.7ex]
 -&S_1^-(u-i)&S_2^+(u-\sfrac{i}{2})&U_{+1}(u)\\[0.7ex]
 +&S_3^-(u-i)&S_2^+(u-\sfrac{i}{2})&U_{+1}(u)\\[0.7ex]
 +&S_3^+(u)  &S_4^-(u-\sfrac{i}{2})&U_{+1}(u)\\[0.7ex]
 +&S_5^-(u)  &S_4^+(u+\sfrac{i}{2})&U_{-1}(u)\\[0.7ex]
 +&S_5^+(u+i)&S_6^-(u+\sfrac{i}{2})&U_{-1}(u)\\[0.7ex]
 -&S_7^+(u+i)&S_6^-(u+\sfrac{i}{2})&U_{-1}(u)\\[0.7ex]
 -&S_7^-(u)  &                     &U_{-1}(u)
\end{array}
\]
with the scattering and potential terms
\[\label{eq:CompFcns.}
S_j^{\pm}(u)=\prod_{p=1}^{K_j} \frac{u-u^{(j)}_p\pm
 \sfrac{i}{2}}{u-u^{(j)}_p\mp \sfrac{i}{2}}\qquad
\mbox{and}\qquad
U_s(u)=\(\frac{u+ \sfrac{i}{2}s}{u}\)^L.
\]
{}From a transfer matrix one can read off the Bethe equations:
The equation for a root $u_{p}^{(j)}$ is equivalent
to the cancellation of poles in $T_{\rep{4|4}}(u)$
(except for the trivial $L$-fold pole at $u=0$)
from each $S_j^{\pm}$ entering in two consecutive terms.
The resulting Bethe equations are spelled out in
\eqref{eq:Beauty.GaugeBethe}.

In the thermodynamic limit, the leading order of
the transfer matrix is determined through
the quasi-momenta introduced in \eqref{QUASIP}
\[\label{alskfjasd}
T_{\rep{4|4}}(u)=\sum_{k=1}^8 \sgrad_k \exp(ip_k(u))+\cdots\,.
\]

Transfer matrices exist for any representation of the symmetry algebra.
Those in oscillator representations, e.g., are generated by
the characteristic determinant, see \appref{sec:Spectral}.

\subsection{Dualization of the Transfer Matrix}
\label{sec:Beauty.TransferDual}

The eigenvalue of the transfer matrix \eqref{eq:TMBeauty} is made up of
the scattering function and the potential function \eqref{eq:CompFcns.}.
For these functions the following relations
are derived from the definition of dual roots:
\[\label{eq:dualssterm}
S_u^+(u)
\left(S_v^-(u-\sfrac{i}{2})-S_w^-(u-\sfrac{i}{2})\right)
=
S_{\tilde u}^-(u)
\left(S_w^+(u+\sfrac{i}{2})-S_v^+(u+\sfrac{i}{2})\right)
\]
for $s_u=0$ and
\<\label{eq:dualssuterm}
\earel{}
S_u^+(u)
\left(S_v^-(u-\sfrac{i}{2})U_{t+s_u-1}(u-\sfrac{i}{2}t)
     -S_w^-(u-\sfrac{i}{2})U_{t-s_u-1}(u-\sfrac{i}{2}t)\right)
\nln
\eq
S_{\tilde u}^-(u)
\left(S_w^+(u+\sfrac{i}{2})U_{t+s_u+1}(u-\sfrac{i}{2}t)
     -S_v^+(u+\sfrac{i}{2})U_{t-s_u+1}(u-\sfrac{i}{2}t)\right)
\>
for $s_u\ne 0$.
Here $S_{\tilde u}^{\pm}(u)$ denotes the scattering function
for the dual roots $\{\tu_j\}$ and $t$ is an arbitrary parameter.
These relations provide us with the duality transformation
among transfer matrices.
For example, one can transform
$T_{\rep{4|4}}(u)$ in \eqref{eq:TMBeauty}
into
\[\label{eq:TMBeast}
\begin{array}[b]{r@{\,}l@{\;}l@{\;}l}
T_{\rep{4|4}}^{\mathrm{Beast}}(u)=
 -&   S_1^+(u)   &                         &U_{+1}(u)\\[.7ex]
 -&   S_1^-(u-i) &   S_2^+(u-\sfrac{i}{2}) &U_{+1}(u)\\[.7ex]
 -& \tS_3^+(u-i) &   S_2^-(u-\sfrac{3i}{2})&U_{-5}(u)\\[.7ex]
 -& \tS_3^-(u-2i)& \tS_4^+(u-\sfrac{3i}{2})&U_{-1}(u)\\[.7ex]
 +&\dtS_5^-(u-2i)& \tS_4^+(u-\sfrac{3i}{2})&U_{-1}(u)\\[.7ex]
 +&\dtS_5^+(u-i) &\dtS_6^-(u-\sfrac{3i}{2})&U_{-1}(u)\\[.7ex]
 +& \tS_7^-(u-i) &\dtS_6^+(u-\sfrac{i}{2}) &U_{-1}(u)\\[.7ex]
 +& \tS_7^+(u)   &                         &U_{-1}(u)
\end{array}
\]
by following the sequence of duality transformations
as illustrated in \figref{fig:B2B}.
Scattering terms for dualized roots are denoted by $\tS,\dtS$.
Another example is obtained from the transformation
illustrated in \figref{fig:su2sl}.
One can also transform $T_{\rep{4|4}}(u)$ into
\[\label{eq:TMBeauty2}
\begin{array}[b]{r@{\,}l@{\;}l@{\;}l}
T_{\rep{4|4}}^{\mathrm{Beauty'}}(u)=
 +&\tS_1^-(u)  &                        &U_{+1}(u)\\[.7ex]
 +&\tS_1^+(u+i)&\dtS_2^-(u+\sfrac{i}{2})&U_{+1}(u)\\[.7ex]
 -&\tS_3^+(u+i)&\dtS_2^-(u+\sfrac{i}{2})&U_{+1}(u)\\[.7ex]
 -&\tS_3^-(u)  &   S_4^+(u+\sfrac{i}{2})&U_{+1}(u)\\[.7ex]
 -&\tS_5^+(u)  &   S_4^-(u-\sfrac{i}{2})&U_{-1}(u)\\[.7ex]
 -&\tS_5^-(u-i)&\dtS_6^+(u-\sfrac{i}{2})&U_{-1}(u)\\[.7ex]
 +&\tS_7^-(u-i)&\dtS_6^+(u-\sfrac{i}{2})&U_{-1}(u)\\[.7ex]
 +&\tS_7^+(u)  &                        &U_{-1}(u),
\end{array}
\]
which corresponds to the second diagram from the bottom.
The transfer matrix corresponding to the bottom
has the same form as \eqref{eq:TMBeauty}
except that all the spins $s$ of $U_s(u)$ are inverted.%
\footnote{This corresponds to the fact that
both of the top and bottom of Fig.~\ref{fig:su2sl}
are in the ``Beauty{}'' form,
but carry opposite Dynkin indices.}
This one and \eqref{eq:TMBeauty2} are equivalent
in the sense that both of them produce the same Bethe equations.

\section{Characteristic Determinant}
\label{sec:Spectral}

Let us briefly discuss the characteristic determinant of
our spin chain.
The characteristic determinant is part of the Baxter equation,
which is an important equation for quantum integrable models.
We will however focus on a different aspect, namely
that the characteristic determinant can be understood as
a generating operator for transfer matrix eigenvalues
in various representations.
This represents a shortcut to the process of fusion
of the fundamental transfer matrix with itself
\cite{Krichever:1997qd}.
In case of the unitary algebra $\alg{su}(m)$,
the characteristic determinant $\Psi_u$ is of the form
\cite{Krichever:1997qd} (see also \cite{Kuniba:1994na})
\[\label{eq:specgeneric}
\Psi_u\sim
\prod_{k=1}^m
\lrbrk{\exp(i\partial_u)
     -S_{k}^-(u+\cdots)\,S_{k-1}^+(u+\cdots)\,U_{\ldots}(u+\cdots) }
\]
with $S$ and $U$ as in \eqref{eq:CompFcns.}.
The omitted constants will not be of importance here.

\paragraph{Antisymmetric Representations.}

The characteristic determinant can be brought 
to the form of a difference operator
\[\label{eq:DIFFO}
 \Psi_u=\sum_{s=0}^{m}(-1)^s e^{is\partial_u/2}\, T_{\rep{[s]}}(u)\,e^{is\partial_u/2}.
\]
The coefficients are given by the eigenvalues of
transfer matrices $T_{\rep{[s]}}(u)$ in totally antisymmetric
products of the fundamental representation.
The characteristic determinant is thus a generating operator
for transfer matrix eigenvalues.

\paragraph{Symmetric Representations.}

The eigenvalues of transfer matrices in totally
symmetric products of the fundamental representation
$T_{\rep{\{s\}}}(u)$ are also contained in
$\Psi_u$, or more precisely in the inverse
\[
\Psi^{-1}_u=\sum_{s=0}^{\infty} e^{-is\partial_u/2-im/2}\,
T_{\rep{\{\bar s\}}}(u)\,e^{-is\partial_u/2-im/2}.
\]
Here we expand all the factors within $\Psi_u^{-1}$ according to
\[\label{eq:specexpand1}
\lrbrk{\exp(i\partial_u)-F(u)}^{-1}
=
\exp(-i\partial_u)+\exp(-i\partial_u)F(u)\exp(-i\partial_u)+\cdots\,.
\]
in other words we assume $\exp(i\partial_u)$ to be large.
An alternative mode of expansion leads to
the eigenvalues of transfer matrices in totally
symmetric products of the antifundamental representation
$T_{\rep{\{\bar s\}}}(u)$:
\[
\Psi^{-1}_u=(-1)^m\sum_{s=0}^{\infty} e^{is\partial_u/2}\,T_{\rep{\{s\}}}(u)\,e^{is\partial_u/2}.
\]
To obtain this form, we consider $\exp(i\partial_u)$ to be small
\[\label{eq:specexpand2}
\lrbrk{\exp(i\partial_u)-F(u)}^{-1}
=
-F(u)^{-1}
-F(u)^{-1}\exp(i\partial_u)F(u)^{-1}
-\cdots\,,
\]
%

\paragraph{Non-Compact Representations.}

This does not even exhaust the set of representations
contained in $\Psi_u$.
We can choose to expand some of the factors within
$\Psi_u^{-1}$ according to \eqref{eq:specexpand1}
and the others according to \eqref{eq:specexpand2}.
This leads to an expansion
\[
\Psi^{-1}_u=(-1)^l\sum_{s=-\infty}^{\infty} e^{is\partial_u/2}\,
T_{\rep{\{l,s\}}}(u)\,e^{is\partial_u/2}.
\]
in terms of transfer matrix eigenvalues $T_{\rep{\{l,s\}}}(u)$
in some representations $\rep{\{l,s\}}$. Here $l$ is the number
of factors expanded according to \eqref{eq:specexpand2}.
All these representations are infinite-dimensional,
the transfer matrix eigenvalues thus contain infinitely many terms.
This expansion is particularly useful for non-compact forms
$\alg{su}(m,n)$ of the unitary algebra, especially the
the conformal group $\alg{su}(2,2)$. The representations
taken by fields within a conformal field theory
are of the form $\rep{\{2,s\}}$, where $s$ corresponds
to the conformal spin.

\paragraph{Oscillator Representations.}

The representations appearing above can be summarized
as the oscillator representations of $\alg{su}(m)$.
Let us briefly outline the relationship
between oscillator representation and the above expansions of the
characteristic determinant.
One introduces a set of $m$ oscillators $\mathbf{a}^k$ and their
conjugates $\mathbf{a}^\dagger_k$.
The generators of $\alg{su}(m)$ are represented by mixed bilinears
$\mathbf{a}^\dagger_k \mathbf{a}^l$.
The oscillator number $M=\sum_{k=1}^m\mathbf{a}^\dagger_k \mathbf{a}^k$
commutes with $\alg{su}(m)$.
Thus the oscillator representation naturally splits into
various irreducible representations labelled by the
eigenvalues of $M$.
This label corresponds to the parameter $k$ in the above
expansions of $\Psi_u$.
Now we can choose the oscillators to obey commutation
or anticommutation relations.
Fermionic oscillators lead to totally antisymmetric
products of the fundamental representation
and thus to the expansion of $\Psi_u$ in \eqref{eq:DIFFO}.
Conversely, bosonic generators correspond to
the expansions of the inverse characteristic determinant
$\Psi^{-1}_u$.
Here we can choose, for each oscillator $k$, whether
the oscillator vacuum should be annihilated by
$\mathbf{a}^k$ or by $\mathbf{a}^\dagger_k$.%
\footnote{For fermions this choice does not make a difference
in analogy to the unique expansion of
the polynomial factor $\exp(i\partial_u)-F(u)$.}
These two choices correspond to the expansions
\eqref{eq:specexpand1} or \eqref{eq:specexpand2}
of the inverse factors within $\Psi_u^{-1}$.

\paragraph{The Superalgebra $\alg{su}(2,2|4)$.}

For the superalgebra, it is natural to expect
the existence of a similar characteristic determinant.
As the oscillator representation of a supersymmetric
algebra contains bosonic and fermionic oscillators,
we should expect the characteristic determinant to consist
of regular and inverse factors.
Using the knowledge of \eqref{eq:TMBeauty} and of the Bethe equations
for the superalgebra $\alg{su}(2,2|4)$, we suggest the following
form of the characteristic operator
to generalize \eqref{eq:specgeneric}
\<\label{eq:QCURVE}
\Psi_u\eq \phantom{\cdot}
      \lrbrk{\exp(i\partial_u)-S_1^+(u+\sfrac{i}{2})\ \hspace{1.9cm}
      U_{+1}(u+\sfrac{i}{2}) }^{-1} \nl
\cdot \lrbrk{\exp(i\partial_u)-S_1^-(u+\sfrac{i}{2})\ S_2^+(u+i)\
      U_{+1}(u+\sfrac{3i}{2}) }^{-1} \nl
\cdot \lrbrk{\exp(i\partial_u)-S_3^-(u+\sfrac{i}{2})\ S_2^+(u+i)\
      U_{+1}(u+\sfrac{3i}{2}) } \nl
\cdot \lrbrk{\exp(i\partial_u)-S_3^+(u+\sfrac{i}{2})\ S_4^-(u)\hspace{0.8cm}
      U_{+1}(u+\sfrac{i}{2}) } \nl
\cdot \lrbrk{\exp(i\partial_u)-S_5^-(u-\sfrac{i}{2})\ S_4^+(u)\hspace{0.8cm}
      U_{-1}(u-\sfrac{i}{2}) } \nl
\cdot \lrbrk{\exp(i\partial_u)-S_5^+(u-\sfrac{i}{2})\ S_6^-(u-i)\
      U_{-1}(u-\sfrac{3i}{2}) } \nl
\cdot \lrbrk{\exp(i\partial_u)-S_7^+(u-\sfrac{i}{2})\ S_6^-(u-i)\
      U_{-1}(u-\sfrac{3i}{2}) }^{-1} \nl
\cdot \lrbrk{\exp(i\partial_u)-S_7^-(u-\sfrac{i}{2})\ \hspace{1.9cm}
      U_{-1}(u-\sfrac{i}{2}) }^{-1},
\>
see \eqref{eq:CompFcns.} for the definition of $S$ and $U$.

There are various ways of expanding this expression.
We can consider all $\exp(i\partial_u)$ to be large
and obtain
\[\label{eq:QCE}
\Psi_u =\sum_{s=0}^\infty (-1)^s e^{-is\partial_u/2} T_{\rep{[s\}}}(u)\,e^{-is\partial_u/2},
\qquad
\Psi^{-1}_u =\sum_{s=0}^\infty e^{-is\partial_u/2}\, T_{\rep{\{s]}}(u)\,e^{-is\partial_u/2}.
\]
The expansion coefficients are the transfer matrix eigenvalues
in super(anti)symmetric representations;
supersymmetric means that the representation is totally symmetric
w.r.t.~$\alg{su}(4)$ and totally antisymmetric w.r.t.~$\alg{su}(2,2)$
and vice versa for superantisymmetric.
Both towers are infinite because they contain totally symmetric
representations of either $\alg{su}(4)$ or $\alg{su}(2,2)$.
The fundamental transfer matrix \eqref{eq:TMBeauty}
appears at the first level of both towers.
\[
T_{\rep{4|4}}(u)=T_{\rep{[1\}}}(u)=T_{\rep{\{1]}}(u)
\]
Similarly, we can consider all $\exp(i\partial_u)$ to be small
and obtain the conjugate representations
\[
\Psi_u =\sum_{s=0}^\infty (-1)^s e^{is\partial_u/2} T_{\rep{[\bar s\}}}(u)\,e^{is\partial_u/2},
\qquad
\Psi^{-1}_u =\sum_{s=0}^\infty e^{is\partial_u/2}\, T_{\rep{\{\bar s]}}(u)\,e^{is\partial_u/2}.
\]
A further important mode of expansion is
as follows. The first two factors
in \eqref{eq:QCURVE}
should be expanded according to
\eqref{eq:specexpand1},
the last two using \eqref{eq:specexpand2}.
We then obtain transfer matrix elements in infinite-dimensional
representations
\[
\Psi_u=\sum_{s=-\infty}^\infty (-1)^s e^{is\partial_u/2} T_{\rep{(s)}}(u)\,e^{is\partial_u/2}.
\]
The transfer matrix eigenvalue
$T_{\rep{(0)}}(u)$
for $s=0$ turns out
to be the generator of local charges $Q_r$.

\paragraph{Dualization.}

The duality transformation of \secref{sec:Dualize}
can be applied to the characteristic determinant.
Let us consider the product of two consecutive terms
within $\Psi_u$. For these the following two identities hold for
$s_u=0$
\<
\earel{}
\phantom{\cdot}
 \lrbrk{\exp(i\partial_u)-S_u^+(u+\sfrac{i}{2})\ S_v^-(u)\
      F(u+\sfrac{i}{2})}^{-1} \nl
\cdot \lrbrk{\exp(i\partial_u)-S_u^+(u+\sfrac{i}{2})\ S_w^-(u)\
      F(u+\sfrac{i}{2})}
\nln
\eq
\phantom{\cdot}
 \lrbrk{\exp(i\partial_u)-S_{\tu}^-(u-\sfrac{i}{2})\ S_v^+(u)\
      F(u-\sfrac{i}{2})} \nl
\cdot \lrbrk{\exp(i\partial_u)-S_{\tu}^-(u-\sfrac{i}{2})\ S_w^+(u)\
      F(u-\sfrac{i}{2})}^{-1} .
\>
with some arbitrary function $F(u)$
and for $s_u\neq 0$
\<
\earel{}
\phantom{\cdot}
 \lrbrk{\exp(i\partial_u)-S_u^+(u+\sfrac{i}{2})\ S_v^-(u)\
      U_{t-1+s_u}(u-\sfrac{i}{2}t+\sfrac{i}{2})\ F(u+\sfrac{i}{2})}^{-1} \nl
\cdot \lrbrk{\exp(i\partial_u)-S_u^+(u+\sfrac{i}{2})\ S_w^-(u)\
      U_{t-1-s_u}(u-\sfrac{i}{2}t+\sfrac{i}{2})\ F(u+\sfrac{i}{2})}
\nln
\eq
\phantom{\cdot}
 \lrbrk{\exp(i\partial_u)-S_{\tu}^-(u-\sfrac{i}{2})\ S_v^+(u)\
      U_{t+1-s_u}(u-\sfrac{i}{2}t-\sfrac{i}{2})\ F(u-\sfrac{i}{2})} \nl
\cdot \lrbrk{\exp(i\partial_u)-S_{\tu}^-(u-\sfrac{i}{2})\ S_w^+(u)\
      U_{t+1+s_u}(u-\sfrac{i}{2}t-\sfrac{i}{2})\ F(u-\sfrac{i}{2})}^{-1} .
\>
To show the agreement, we have to multiply
the inverse terms to the other side and expand.
The identities \eqref{eq:dualssterm,eq:dualssuterm}
then guarantee the equality.
Dualization of the characteristic determinant thus consists of permuting
one factor with an adjacent inverse factor
according to the above relations.

\section{Anomaly}
\label{sec:anom}

Consider two sets of roots $\{u_p\}_{p=1}^K$ and $\{v_p\}_{p=1}^K$
which are pairwise close to each other: $v_p=u_p+\order{1}$.
It is the situation which we encounter in strings of stacks,
where two roots $u_p, v_p$ belong to the same $p$-th stack,
and the size of the string is large $K=\order{L}$.
In other words, each $v$-root is attached to one of the $u$-roots.
Let us denote the distance between $u$- and $v$-roots by
$\xi_p=v_p-u_p=\order{1}$.
Then the scattering factor for root $v_p$ off the roots $\{u_q\}_{q=1}^K$
\[\label{eq:vu-scatt}
\prod_{q=1}^{K}\frac{v_p-u_q-\frac{i}{2}}{v_p-u_q+\frac{i}{2}}
\]
contains two contributions: from short distances, when
$v_p-u_q={\cal O}(1)$, and from large distances,
when $v_p-u_q={\cal O}(L)$.
Let us divide the above product into two parts, $|p-q|\leq A$
and $|p-q|>A$, where $1\ll A\ll L$.

Consider first the
short-distance part of the product. For sufficiently small $r$, we can
approximate
\<
u_{p+r}\eq u_p+\frac{r}{\rho (u_p/L)}+\order{1/L},
\nln
v_{p+r}\eq u_p+\xi_p+\frac{r}{\rho (u_p/L)}+\order{1/L},
\>
and neglect the $1/L$ correction as far as $|r|\ll L$.
Note that the density of roots $\rho(u)$ at $uL\approx u_p$
is the same for $u_p$ as for $v_p$ by construction.
Then
\[
 \prod_{r=-A}^{A}
 \frac{v_p-u_{p+r}-\frac{i}{2}}{v_p-u_{p+r}+\frac{i}{2}}
 =\prod_{r=-A}^{A}
 \frac{\xi _p-\frac{i}{2}-{r}/{\rho (u_p/L)}}
 {\xi _p+\frac{i}{2}-{r}/{\rho (u_p/L)}}
+\cdots
\]
holds up to ${\cal O}(1/L)$ corrections. Since $A\gg 1$ and the
product quickly converges, we can replace the upper limit of
multiplication by infinity and then find the full anomaly
contribution, similarly to \cite{Daul:1993xz}
\[
  \prod_{r=-A}^{A}
 \frac{v_p-u_{p+r}-\frac{i}{2}}{v_p-u_{p+r}+\frac{i}{2}}
 =
 \frac{\sin\left(\pi \rho(u_p/L)\left(\xi_p-\frac{i}{2}\right)\right)}
 {\sin\left(\pi \rho(u_p/L)\left(\xi_p+\frac{i}{2}\right)\right)}
+\cdots\,.
\]

If $|q-p|\gg 1$, we can neglect ${\cal O}(1)$ terms in $v_p-u_q$ and
approximate $v_p$ by $u_p$. Hence, the large-distance part of the
product in \eqref{eq:vu-scatt} is
\[
  \prod_{|q-p|>A}\frac{v_p-u_q-\frac{i}{2}}{v_p-u_q+\frac{i}{2}}
 =\exp\bigbrk{i\resolvsl(u_p/L)+\cdots},
\]
where
\[ \resolvsl(u)=\pint \frac{dv\,\rho(v)}{v-u}
\]
is the principal value of the $u$-resolvent
(which coincides with the $v$-resolvent).

Finally we get
\[
 \label{ANOMP}
 \prod_{q=1}^K\frac{v_p-u_q-\frac{i}{2}}{v_p-u_q+\frac{i}{2}}
 =
 \exp\bigbrk{i H_{vu}(u_p/L)+i\resolvsl(u_p/L) +\cdots},
\]
where we introduced the \emph{anomaly} term stemming from the $\xi$
shift between the roots $v$ and $u$
\[\label{asdjfads}
 H_{vu}(u)=
\frac{1}{i}\log\frac{\sin\bigbrk{\pi \rho(u)(\xi(u)-\frac{i}{2})}}
 {\sin\bigbrk{\pi \rho(u)(\xi(u)+\frac{i}{2})}}\, .
\]
The function $\xi(u)\approx \xi_p$ represents the $\order{1}$ separation
of the strings of $u$- and $v$-roots at position $uL\approx u_p$.
Note that if we considered the scattering of roots
within the same flavor then $v_p=u_p$, i.e.~$\xi_p=0$ for any $p$, and we
would get no anomaly term in \eqref{ANOMP} in the leading order,
$H_{uu}(u)=0$. The formula \eqref{asdjfads} does not directly
generalize to $\xi(u)=0$ due to omission of the term $p=q$
for alike roots in the Bethe equations.

Now we can verify that the scattering factor between two stacks
discussed in \secref{sec:Stacks.Stacks} is anomaly free.
The scattering factor \eqref{eq:stackint} can be decomposed as
follows
\<
s_{kl,kl}(u,u')\eq \prod_{j=k}^{l-1}
\frac{u^{(j)}-u'^{(j)}+\frac{i}{2}M_{j,j}}
     {u^{(j)}-u'^{(j)}-\frac{i}{2}M_{j,j}}
\nl
\times
\prod_{j=k}^{l-1} \prod_{j'=j+1}^{l-1} \lrbrk{
\frac{u^{(j)}-u^{\prime(j')}+\frac{i}{2}M_{j,j'}}
     {u^{(j)}-u^{\prime(j')}-\frac{i}{2}M_{j,j'}}\,
\frac{u^{(j')}-u^{\prime(j)}+\frac{i}{2}M_{j,j'}}
     {u^{(j')}-u^{\prime(j)}-\frac{i}{2}M_{j,j'}}} \,.
\>
Terms in the single product do not yield any anomaly,
as is explained above.
On the other hand, each term in the double products yields an anomaly.
However, by observing the odd property of the anomaly term
\[
H_{vu}(u)=-H_{uv}(u),
\]
we see that anomalies completely cancel out in total,
as far as dealing with Bethe equations corresponding to
a symmetric Cartan matrix $M_{j,j'}=M_{j',j}$.

\bibliography{bksz}
\bibliographystyle{nb}

\end{document}